\documentclass{aastex62}

\accepted{September 10, 2019}
\submitjournal{ApJ}

\shorttitle{ALMA Observations of V900 Mon}
\shortauthors{Takami et al.}
\usepackage{color}
\usepackage{soul,xcolor}
\setstcolor{red}

\begin{document}

\title{An ALMA Study of the FU-Ori Type Object V900 Mon: Implications for the Progenitor
%\footnote{Should we acknowledge ALMA?}
}

\correspondingauthor{Michihiro Takami}
\email{hiro@asiaa.sinica.edu.tw}

\author{Michihiro Takami}
\affil{Institute of Astronomy and Astrophysics, Academia Sinica, 
11F of Astronomy-Mathematics Building, AS/NTU
No.1, Sec. 4, Roosevelt Rd, Taipei 10617, Taiwan, R.O.C.}

\author{Tsu-Sheng Chen}
\affil{Institute of Astronomy and Astrophysics, Academia Sinica, 
11F of Astronomy-Mathematics Building, AS/NTU
No.1, Sec. 4, Roosevelt Rd, Taipei 10617, Taiwan, R.O.C.}

\author{Hauyu Baobab Liu}
\affil{Institute of Astronomy and Astrophysics, Academia Sinica, 
11F of Astronomy-Mathematics Building, AS/NTU
No.1, Sec. 4, Roosevelt Rd, Taipei 10617, Taiwan, R.O.C.}

\author{Naomi Hirano}
\affil{Institute of Astronomy and Astrophysics, Academia Sinica, 
11F of Astronomy-Mathematics Building, AS/NTU
No.1, Sec. 4, Roosevelt Rd, Taipei 10617, Taiwan, R.O.C.}

\author{\'Agnes K\'osp\'al}
\affil{Konkoly Observatory, Research Centre for Astronomy and Earth Sciences, Hungarian Academy of Sciences, Konkoly-Thege Mikl\'os \'ut 15-17, 1121 Budapest, Hungary}
\affil{Max Planck Institute for Astronomy, K\"onigstuhl 17, D-69117 Heidelberg, Germany}

\author{P\'eter \'Abrah\'am}
\affil{Konkoly Observatory, Research Centre for Astronomy and Earth Sciences, Hungarian Academy of Sciences, Konkoly-Thege Mikl\'os \'ut 15-17, 1121 Budapest, Hungary}

\author{Eduard I. Vorobyov}
\affil{Research Institute of Physics, Southern Federal University, Roston-on-Don 344090, Russia}
\affil{Department of Astrophysics, University of Vienna, Vienna 1080, Austria}

\author{Fernando Cruz-S\'aenz de Miera}
\affil{Konkoly Observatory, Research Centre for Astronomy and Earth Sciences, Hungarian Academy of Sciences, Konkoly-Thege Mikl\'os \'ut 15-17, 1121 Budapest, Hungary}

\author{Timea Csengeri}
\affil{Max-Planck-Institut f\"ur Radioastronomie, Auf dem H\"ugel 69, D-53121 Bonn, Germany}

\author{Joel Green}
\affil{Space Telescope Science Institute, 3700 San Martin Dr., Baltimore, MD 21218, USA}

\author{Michiel Hogerheijde}
\affil{Leiden Observatory, Leiden University \&
Anton Pannekoek Institute for Astronomy, University of Amsterdam,
PO Box 9513, 2300 RA, Leiden, The Netherlands}

\author{Tien-Hao Hsieh}
\affil{Institute of Astronomy and Astrophysics, Academia Sinica, 
11F of Astronomy-Mathematics Building, AS/NTU
No.1, Sec. 4, Roosevelt Rd, Taipei 10617, Taiwan, R.O.C.}

\author{Jennifer L. Karr}
\affil{Institute of Astronomy and Astrophysics, Academia Sinica, 
11F of Astronomy-Mathematics Building, AS/NTU
No.1, Sec. 4, Roosevelt Rd, Taipei 10617, Taiwan, R.O.C.}

\author{Ruobing Dong}
\affil{Department of Physics \& Astronomy, University of Victoria, Victoria, BC, V8P 1A1, Canada}
\affil{Institute of Astronomy and Astrophysics, Academia Sinica, 
11F of Astronomy-Mathematics Building, AS/NTU
No.1, Sec. 4, Roosevelt Rd, Taipei 10617, Taiwan, R.O.C.}

\author{Alfonso Trejo}
\affil{Institute of Astronomy and Astrophysics, Academia Sinica, 
11F of Astronomy-Mathematics Building, AS/NTU
No.1, Sec. 4, Roosevelt Rd, Taipei 10617, Taiwan, R.O.C.}

\author{Lei Chen}
\affil{Konkoly Observatory, Research Centre for Astronomy and Earth Sciences, Hungarian Academy of Sciences, Konkoly-Thege Mikl\'os \'ut 15-17, 1121 Budapest, Hungary}

%% Note that the \and command from previous versions of AASTeX is now
%% depreciated in this version as it is no longer necessary. AASTeX 
%% automatically takes care of all commas and "and"s between authors names.

%% AASTeX 6.2 has the new \collaboration and \nocollaboration commands to
%% provide the collaboration status of a group of authors. These commands 
%% can be used either before or after the list of corresponding authors. The
%% argument for \collaboration is the collaboration identifier. Authors are
%% encouraged to surround collaboration identifiers with ()s. The 
%% \nocollaboration command takes no argument and exists to indicate that
%% the nearby authors are not part of surrounding collaborations.

%% Mark off the abstract in the ``abstract'' environment. 
\begin{abstract}
We present ALMA observations of $^{12}$CO, $^{13}$CO, and C$^{18}$O $J$=2--1 lines and the 230 GHz continuum for the FU Ori-type object (FUor) V900 Mon ($d$$\sim$1.5 kpc), for which the accretion burst was triggered between 1953 and 2009. We identified CO emission associated with a molecular bipolar outflow extending up to a $\sim$10$^4$ au scale and a rotating molecular envelope extending over $>$10$^4$ au. The interaction with the hot energetic FUor wind, which was observed using optical spectroscopy, appears limited to a region within $\sim$400 au of the star. The envelope mass and the collimation of the extended CO outflow suggest that the progenitor of this FUor is a low-mass Class I young stellar object (YSO). These parameters for V900 Mon, another FUor, and a few FUor-like stars are consistent with the idea that FUor outbursts are associated with normal YSOs.
The continuum emission is marginally resolved in our observations with a 0\farcs2$\times$0\farcs15 ($\sim$300$\times$225 au) beam, and a Gaussian model provides a deconvolved FWHM of $\sim$90 au. The emission is presumably associated with a dusty circumstellar disk, plus a possible contribution from a wind or a wind cavity close to the star. The warm compact nature of the disk continuum emission could be explained with viscous heating of the disk, while gravitational fragmentation in the outer disk and/or a combination of grain growth and their inward drift may also contribute to its compact nature.
\end{abstract}

%% Keywords should appear after the \end{abstract} command. 
%% See the online documentation for the full list of available subject
%% keywords and the rules for their use.

%% From the front matter, we move on to the body of the paper.
%% Sections are demarcated by \section and \subsection, respectively.
%% Observe the use of the LaTeX \label
%% command after the \subsection to give a symbolic KEY to the
%% subsection for cross-referencing in a \ref command.
%% You can use LaTeX's \ref and \label commands to keep track of
%% cross-references to sections, equations, tables, and figures.
%% That way, if you change the order of any elements, LaTeX will
%% automatically renumber them.
%%
%% We recommend that authors also use the natbib \citep
%% and \citet commands to identify citations.  The citations are
%% tied to the reference list via symbolic KEYs. The KEY corresponds
%% to the KEY in the \bibitem in the reference list below. 

%%%%%%%%%%%%%%%%%%%%%%%%%%%%%%%%%%%%%%%%
%%%%%%%%%%%%%%%%%%%%%%%%%%%%%%%%%%%%%%%%
%%%%%%%%%%%%%%%%%%%%%%%%%%%%%%%%%%%%%%%%
%% Section : Introduction
%%%%%%%%%%%%%%%%%%%%%%%%%%%%%%%%%%%%%%%%
%%%%%%%%%%%%%%%%%%%%%%%%%%%%%%%%%%%%%%%%
%%%%%%%%%%%%%%%%%%%%%%%%%%%%%%%%%%%%%%%%

\section{Introduction} \label{sec:intro}

Most of the stars in our Galaxy have masses below a few solar masses. We do not understand the physical mechanism by which these stars (``low-mass" stars) accrete their masses well.

The evolution of low-mass young stellar objects (YSOs) is characterized and classified using their infrared spectral energy distribution (SED) \citep[Class 0$\rightarrow$I$\rightarrow$II$\rightarrow$III;][]{Stahler05}. \citet{Muzerolle98_IR} measured the mass accretion rates for a sample of Class I-II YSOs using the Br $\gamma$ line, and showed that steady mass accretion can explain only a fraction of their final stellar masses.  This issue is also corroborated by the facts below. At the pre-main sequence phase of their evolution (``Class II-III''), in which the circumstellar gas+dust envelope has already been dissipated, the masses of the associated circumstellar disks are significantly smaller than the stars \citep[e.g.,][]{Williams11}, indicating that the stellar masses have been developed primarily during the Class 0-I phases. %At these younger stages, for which the total luminosity of each system is dominated by the accretion luminosity.
It has been suggested that their protostellar luminosities
%the accretion luminosity of such young systems, which is equivalent to the total luminosity,
tend to be significantly lower than theoretical predictions for steady mass accretion, e.g., by a factor of 10-10$^3$ \citep[the ``luminosity problem''; see e.g.,][]{Kenyon90,Dunham14,Audard14}.

%In contrast, \citet{Dunham12} analyzed a number of their numerical simulations and concluded that the contribution of the accretion outbursts is not significant for the final mass of the star.

A key phenomenon that may solve the above issues is episodic mass accretion, observed in some YSOs as a sudden increase of flux at optical and near-infrared (IR) wavelengths. The above trends are explained if Class 0-I YSOs are associated with accretion outbursts whose periods are significantly shorter than the time scale of these evolutionary phases (and therefore with a small chance of observation) but which are responsible for a significant fraction of the final stellar masses \citep[e.g.,][]{Kenyon90,Muzerolle98_IR,Calvet00}. 
%In this context, the YSOs with accretion outbursts may hold a key to understand the physical mechanism for many stars to accrete their masses.

The FU Orionis objects (hereafter FUors) are a class of YSOs which undergo the most active and violent accretion outbursts during which the accretion rate rapidly increases by a factor of 100-1000, and remains high for several decades or more. Such outbursts have been observed toward about 10 stars to date. Another dozen YSOs exhibit optical or near-IR spectra similar to FUors, distinct from many other YSOs, but outbursts have never been observed. Their spectra suggest disk accretion with high accretion rates similar to FUors. These are classified as FUor candidates or FUor-like objects. See \citet{Audard14} for a review for FUors and FUor-like objects. Their optical and near-IR spectra indicate that the optical and near-IR emission from these objects is dominated by a warm disk photosphere \citep[][for reviews]{Hartmann96,Audard14}. This is in contrast to many other normal Class I-II YSOs, whose optical and near-IR emission is dominated by the star and featureless dust continuum \citep[e.g.,][]{Greene96,Doppmann05,Connelley10}. 
%Many FUors are known to host a massive circumstellar gas+dust envelope comparable to normal Class I YSOs \citep[e.g.,][]{Sandell01,Millan-Gabet06,Kospal17b,Kospal17c,Feher17}, consistent with the above hypothesis that many Class I YSOs experience such accretion outbursts.
A sudden increase in the accretion rate heats up the inner disk ($r$$<$1 au) of the FUors, observable as continuum emission at optical and IR wavelengths.

The triggering mechanism of FUor outbursts is not clear, despite numerous observations at a variety of wavelengths, theoretical work, and numerical simulations. The proposed mechanisms include: (1) gravitational/thermal/magneto-rotational instabilities in the disk, or (2) the perturbation of the disk by an external body \citep[see][for a review]{Audard14}. 
%{\bf The instabilities in the inner disk could be triggered via mass accumulation from a gravitationally fragmented outer disk  \citep[r$>$50 au; e.g.,][]{Vorobyov05,Vorobyov10,Vorobyov15b,Machida11} or an infalling envelope \citep[see][for a review]{Hartmann96}. }
%An infalling envelope or gravitational instability in the outer disk ($r$$\gg$1 au) may also {\bf trigger} the outbursts \citep[e.g.,][]{Vorobyov05,Vorobyov10,Vorobyov15b,Machida11} {\bf (EV:Infall of dense gaseous clumps formed through gravitational fragmentation occurring in the outer disk (r$>$50 au) and sustained by an infalling envelope  )}. 
%
\citet{Liu16,Takami18} have revealed a variety of distributions in the near-IR scattered light, that can be attributed to gravitationally unstable disks and trails of clump ejections in such disks. These results support gravitational instabilities in disks as a triggering mechanism of the outbursts, as demonstrated with some hydrodynamical simulations \citep[e.g.,][]{Vorobyov15b,Dong16,Zhao18}. \citet{Dunham12} carried out radiative transfer calculations for a number of hydrodynamical simulations with gravitationally unstable disks, and demonstrated that this physical mechanism can also solve the ``luminosity problem" described above. According to their simulations, accretion bursts contribute to only up to 35 \% of the final stellar mass, in contrast to the previous arguments that FUor outbursts are essential for low-mass protostellar evolution \citep[e.g.,][]{Muzerolle98_IR,Calvet00}.

However, another key question still remains: are FUor outbursts associated with {most} low-mass YSOs? In other words, are FUors peculiar YSOs, or is this a specific phase for the evolution of {most} low-mass YSOs?
As with normal YSOs, many of these observations show extended molecular outflows, envelopes and disks associated with FUors and FUor-like objects.
Infrared to Millimeter emission and mid-IR silicate absorption indicate that FUors are associated with an envelope similar to those of Class 0-I YSOs \citep[e.g.,][]{Sandell01,Green06,Green13,Quanz07,Gramajo14,Kospal17b,Feher17}. Mid-IR spectroscopy shows that some FUors are associated with silicate absorption like Class-I YSOs, while other objects are associated with emission like Class II YSOs \citep{Green06,Quanz07}. \citet{Quanz07} attributed this trend to an evolution of the FUors similar to the Class 0-I to Class II transitions of normal YSOs.

Thanks to a combination of high-angular resolution and sensitivity, the Atacama Large Millimeter/submillimeter Array (ALMA) has recently begun to provide detailed information about their cold ($\ll$100 K) extended molecular outflows, envelopes and disks, which is potentially powerful for further investigating whether FUor outbursts occur in {most} low-mass YSOs. The FUors and FUor-like objects observed using ALMA include FU Ori \citep[FUor;][]{Hales15}, V883 Ori \citep[FUor-like;][]{Cieza16, Ruiz17_V883Ori}, V346 Nor \citep[FUor;][]{Kospal17b}, V2775 Ori \citep[FUor-like;][]{Zurlo17}, HBC 494 \citep[FUor-like;][]{Ruiz17_HBC494}, and L1551 IRS 5 
%(FUor-like, Cruz-S\'aenz de Miera et al., $submitted$).
\citep[FUor-like;][]{Cruz19}.

In this paper we present ALMA observations of 230 GHz ($\lambda$=1.3 mm) continuum and $^{12}$CO, $^{13}$CO, and C$^{18}$O $J$=2--1 lines for the FUor V900 Mon (2MASS 06572222-0823176). V900 Mon has undergone a major increase in brightness between 1953 and 2009, by at least 4 mag~in the optical \citep{Thommes11,Reipurth12}, and it may still be brightening \citep{Varricatt15}. The luminosity ($\sim$200 $L_\sun$%
\footnote{\citet{Reipurth12} measured 106 $L_\sun$ adopting a distance of 1.1 kpc. We scale this luminosity to the recent measurement of the distance based on Gaia DR2 (1.5$_{-0.2}^{+0.3}$ kpc).}) 
and the optical and near-IR spectra observed during the outburst are similar to FUors \citep{Reipurth12,Connelley18}. The star has a much higher reddening ($A_V \sim$13 mag) than many other FUors ($<$5 for many cases) but comparable to FUor-like objects \citep[$A_V$=2-40 mag][]{Audard14}. While the infrared SED suggests the presence of an envelope \citep{Reipurth12,Gramajo14}, its mass may be modest. \citet{Kospal17b} conducted single-dish observations of $^{12}$CO $J$=3-2, $^{12}$CO $J$=4-3, and $^{13}$CO $J$=3-2 with a 15\arcsec-19\arcsec~resolution. Unlike several other FUors, these authors did not detect these CO emissions at the stellar position, providing an upper mass limit for the envelope of 0.1 $M_\sun$. This fact is corroborated by a mass estimate using a detailed model of the SED by \citet{Gramajo14}. The envelope mass derived by these authors is 0.027 $M_\sun$, one of the lowest among the 23 FUors and FUor-like objects in their study (0-1.5 $M_\sun$).

Using ALMA, we were able to detect emission associated with a molecular bipolar outflow, a rotating molecular envelope and a probable disk associated with V900 Mon. The rest of the paper is organized as follows. In Section \ref{sec:obs} we describe the observations. In Section \ref{sec:results} we show various images for continuum and CO emission, and CO line profiles for selected regions. In Section \ref{sec:discussion} we discuss the molecular bipolar outflow, envelope, disk and interaction with a hot ($\gg$1000 K) energetic FUor wind\footnote{In this paper, we use the words ``outflow" and ``wind" following the literature in this research field. Throughout, an ``outflow" in the paper implies extended ($\gg$100 AU) molecular outflows primarily observed in millimeter CO emission. The presence of millimeter CO emission indicates a low temperature for the gas ($<$50 K). As described later, these outflows result from the interaction of a collimated jet or a wide-angled wind with the surrounding gas \citep[e.g.,][]{Lee02,Arce07}. The word ``wind" is used for any other gas outflowing directly from the star or the inner disk, usually poorly collimated. (We use the word ``jet" for a collimated flow.)} observed in optical spectra toward the star. In Section \ref{sec:uni} we discuss whether FUor outbursts are associated with normal YSOs. In Section \ref{sec:conclusions} we give our conclusions. Throughout the paper we adopt a distance of 1.5 kpc \citep[1.5$_{-0.2}^{+0.3}$ kpc including the uncertainty,][]{GaiaDR2}.

%%%%%%%%%%%%%%%%%%%%%%%%%%%%%%%%%%%%%%%%
%%%%%%%%%%%%%%%%%%%%%%%%%%%%%%%%%%%%%%%%
%%%%%%%%%%%%%%%%%%%%%%%%%%%%%%%%%%%%%%%%
%% Section : Observations and Data Reduction
%%%%%%%%%%%%%%%%%%%%%%%%%%%%%%%%%%%%%%%%
%%%%%%%%%%%%%%%%%%%%%%%%%%%%%%%%%%%%%%%%
%%%%%%%%%%%%%%%%%%%%%%%%%%%%%%%%%%%%%%%%

\section{Observations and Data Reduction} \label{sec:obs}

We observed V900 Mon with the ALMA 12 m array on 2017 April 20 and July 27, and with the ACA on 2016 October 11 (Project code: 2016.1.00209.S, PI: Michihiro Takami). The pointing and phase referencing center is R.A. (J2000) = 06:57:22.224, and Decl. (J2000) = --08:23:17.64.
The spectral setup of all of our observations is identical. 
%\textcolor{red}{\bf (or ``The spectral setups of all of our observations are identical.")}
There were two 1.875 GHz wide spectral windows (1.3 km s$^{-1}$ velocity resolution) centered at 216.9 and 232.2 GHz; two 58.6 MHz wide spectral windows to cover $^{13}$CO $J$=2--1 and C$^{18}$O $J$=2--1 at a 0.083 km s$^{-1}$ velocity resolution; and one 58.6 MHz wide spectral window to cover $^{12}$CO $J$=2--1 at a 0.040 km s$^{-1}$ velocity resolution. 
These spectral windows tracked the systemic velocity $V_{\mathrm LSR}$$\sim$13.5 km s$^{-1}$.

These observations covered {\it uv} distance ranges of 20-3500 m and 9.3-46 m, respectively. These configurations provided a largest recoverable scale of 29 arcsec. Other details of the observations are summarized in Table  \ref{tab:obs}.

  %%%%%%%%%%%%%%%%%%%%%%%%%%%%%%%%%%%%%%%%
  %% Table : Summary of the observations
  %%%%%%%%%%%%%%%%%%%%%%%%%%%%%%%%%%%%%%%%

\begin{deluxetable*}{ c c c c c c c c}
\tabletypesize{\footnotesize}
  \tablecaption{Summary of the Observations   \label{tab:obs}}
  \tablehead{
  \colhead{Epoch}  & \colhead{Date}   &   \colhead{Array} &   \colhead{PWV} &    \colhead{Flux/} & \colhead{Passband/}   & \colhead{Gain calibrator flux\tablenotemark{b}}    & \colhead{Time on source} \\  
  (\#)    &  UTC (YYYYMMDD) &  & {(mm)}  & \colhead{Passband calibrators\tablenotemark{b}} &  \colhead{Gain calibrator\tablenotemark{b}} & (Jy) & (minutes)
  }
  \startdata
    1   &   20161011    &  ACA    	& {---\tablenotemark{a}}		& J0522-3627/J0854+2006  &  J0725-0054  &   4.3/1.9  & 4.4 \\
    2   &   20170420    &  ALMA  & {2.7}			& J0656-0323/J0510+1800  &  J0656-0323  &   2.3/0.56 & 3.0 \\
    3   &   20170727    &  ALMA  & {0.5}			& J0522-3627/J0510+1800  &  J0654-0544  &   1.1/0.85  &   9.1 \\
  \enddata
  \tablenotetext{a}{Not recorded.}
  \tablenotetext{b}{Measured at 224 GHz.}
\end{deluxetable*}

The data were manually calibrated and phase self-calibrated using the CASA software package \citep{McMullin07} version 5.4. {The self calibration increased the signal-to-noise ratio only marginally.}
We fitted the continuum baselines using the CASA task {\tt uvcontsub}, and then jointly imaged all continuum data using the CASA task {\tt tclean}. Table \ref{tab:resolutions} summarizes the frequencies and the beams of the continuum and CO lines.
The Briggs Robust=0 weighted continuum image taken with ALMA achieved an 55 $\mu$Jy beam$^{-1}$ root-mean-square (RMS) noise level.

  %%%%%%%%%%%%%%%%%%%%%%%%%%%%%%%%%%%%%%%%
  %% Table : Frequencies and Beam
  %%%%%%%%%%%%%%%%%%%%%%%%%%%%%%%%%%%%%%%%
  
\begin{deluxetable*}{ccccccccccc}
\tablecaption{Frequencies and Angular Resolutions \label{tab:resolutions}}
\tablecolumns{9}
%\tablenum{1}
\tablewidth{0pt}
\tablehead{
\colhead{Data} &
\colhead{Frequency} &
\multicolumn{2}{c}{Beam} &
%\colhead{Beam} &
\multicolumn{2}{c}{Uncertainty\tablenotemark{a}} &
\multicolumn{5}{c}{Convolution} \\
%\colhead{Convolution FWHM} &
%\colhead{} &
%\colhead{} &
%\colhead{} &
%\colhead{}\\
\colhead{} &
\colhead{(GHz)} &
\colhead{Size} &
\colhead{P.A.}  &
\colhead{(K)} &
\colhead{(mJy beam$^{-1}$)} &
\multicolumn{2}{c}{Convolution FWHM} &
\colhead{Final}  &
\multicolumn{2}{c}{Uncertainty\tablenotemark{a}} \\
\colhead{} &
\colhead{} &
\colhead{} &
\colhead{} &
\colhead{} &
\colhead{} &
\colhead{(Pixel)} &
\colhead{(Arcsec)} &
\colhead{Resolution}  &
\colhead{(K)} &
\colhead{(mJy beam$^{-1}$)} \\
}
\startdata
Continuum	& 224.5416	& 0\farcs20$\times$0\farcs15	& --72\fdg3
	& 0.037	& {0.055}
	& --- & --- & --- &  --- & ---	\\
$^{12}$CO $J$=2--1	& 230.5380		& 0\farcs19$\times$0\farcs14	& --71\fdg3
	& {4.0}		& {5.2}
	& 5		& 0\farcs22	& 0\farcs29$\times$0\farcs26 	& 2.1		& {7.8} 	\\
$^{13}$CO $J$=2--1	& 220.3987		& 0\farcs19$\times$0\farcs14	& --71\fdg3
	& {4.1}		& {4.9}
	& 8		& 0\farcs36	& 0\farcs41$\times$0\farcs39 	& 1.4	 	& {10}	\\
C$^{18}$O $J$=2--1	& 219.5604		& 0\farcs19$\times$0\farcs14	& --72\fdg4
	& {3.2}		& {3.8}
	& 11		& 0\farcs47	& 0\farcs51$\times$0\farcs49 	& 0.8		& {9.3}	\\
\enddata
\tablenotetext{a}{Excluding the absolute flux uncertainty of 10 \% for the ALMA Band 6 observations. The values for the lines are calculated for a 0.2 km s$^{-1}$ channel.}
\end{deluxetable*}

The data cubes for the CO lines were made with spatial and velocity pixel sampling of 0\farcs02 and 0.2 km s$^{-1}$, respectively, and were analyzed using python ({\tt numpy}, {\tt scipy}, {\tt matplotlib}).
%We spatially convolved them using a Gaussian with a FWHM of 0\farcs2--0\farcs5 to investigate the intensity distributions and kinematics with optimum angular resolutions and {\bf signal-to-noise} (see Table \ref{tab:convolution}). 
To increase signal-to-noise {of the extended emission}, we spatially convolved each velocity channel map using a two-dimensional Gaussian. In Table \ref{tab:resolutions} we also summarize the parameters of the convolution and the final angular resolutions.
The absence of the Total Power observations probably causes significant missing fluxes ($\gtrsim$20 \%) for these line observations. 
%\textcolor{red}{\bf (AK: Why do you think that we miss CO flux? This object is at 1.5 kpc, and the maximum recoverable scale for the 7m observations is 29 arcsec. That's 43500 au at the distance of V900 Mon! Do you really expect CO emission at such a large spatial scale? If yes, that's more likely coming from the star forming region, and not immediately associated with the FUor. How did you get the 20 \% estimate?)}

%%%%%%%%%%%%%%%%%%%%%%%%%%%%%%%%%%%%%%%%
%%%%%%%%%%%%%%%%%%%%%%%%%%%%%%%%%%%%%%%%
%%%%%%%%%%%%%%%%%%%%%%%%%%%%%%%%%%%%%%%%
%% Section : Results
%%%%%%%%%%%%%%%%%%%%%%%%%%%%%%%%%%%%%%%%
%%%%%%%%%%%%%%%%%%%%%%%%%%%%%%%%%%%%%%%%
%%%%%%%%%%%%%%%%%%%%%%%%%%%%%%%%%%%%%%%%

\section{Results} \label{sec:results}

We present the continuum, the velocity channel maps of the CO lines in a wide field of view (FOV, 13\arcsec$\times$13\arcsec), those close to the continuum source (a 4\arcsec$\times$4\arcsec~FOV), and the line profiles in Section
\ref{sec:results:continuum},
\ref{sec:results:vcm1},
\ref{sec:results:vcm2}, and
\ref{sec:results:profiles}, respectively.

  %%%%%%%%%%%%%%%%%%%%%%%%%%%%%%%%%%%%%%%%
  %%%%%%%%%%%%%%%%%%%%%%%%%%%%%%%%%%%%%%%%
  %% Dust Continuum
  %%%%%%%%%%%%%%%%%%%%%%%%%%%%%%%%%%%%%%%%
  %%%%%%%%%%%%%%%%%%%%%%%%%%%%%%%%%%%%%%%%

\subsection{Continuum} \label{sec:results:continuum}

\begin{figure}[ht!]
\epsscale{0.5}
\plotone{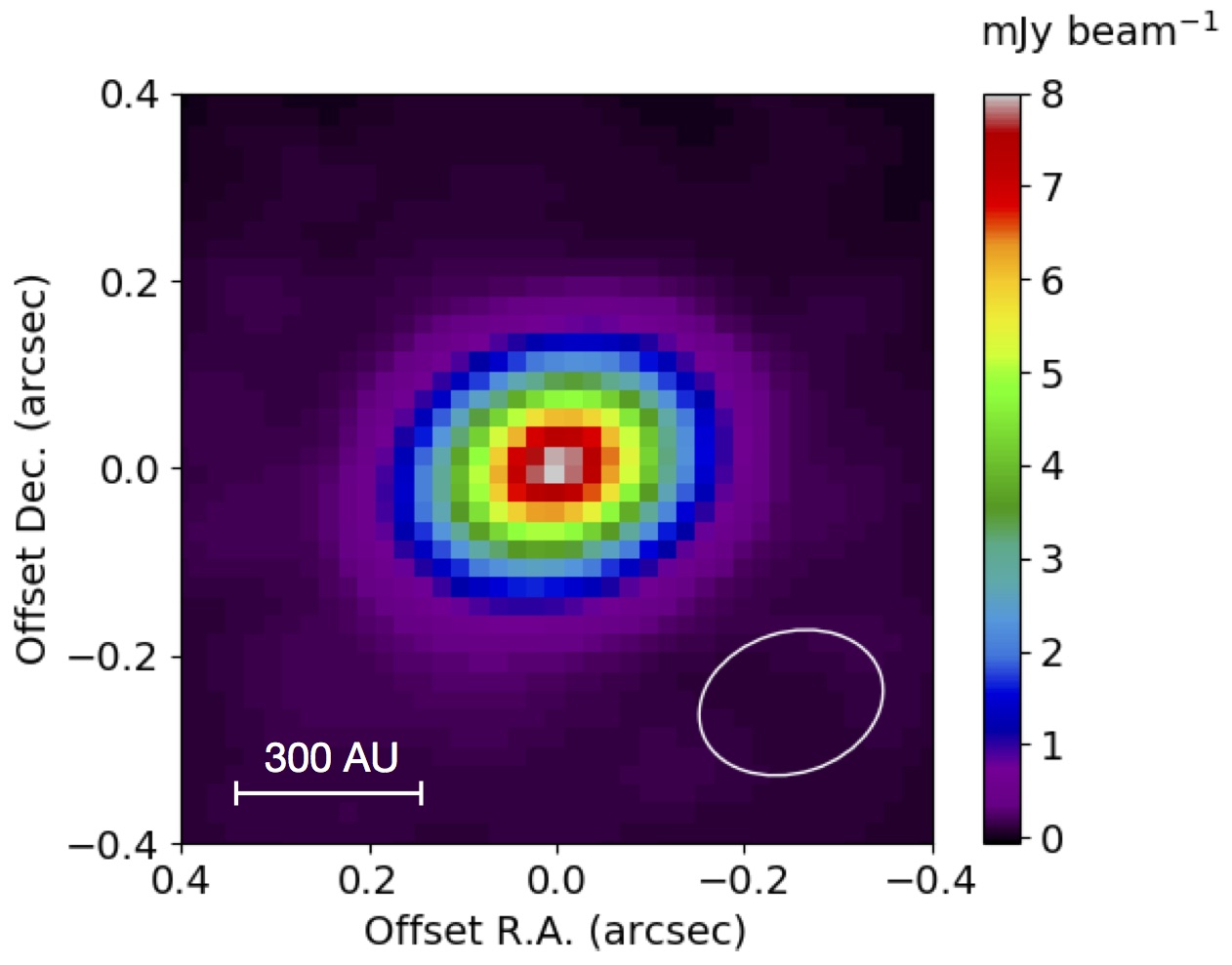}
\caption{The 224.5 GHz continuum image of V900 Mon. The spatial offset is shown from the centroid position measured using Gaussian fitting  (06:57:22.221 $-$8:23:17.67). 
%This position is also shown using the white cross.
The white ellipse shows the beam size of the observations.
\label{fig:continuum}}
\end{figure}

Figure \ref{fig:continuum} shows the our 1.3 mm continuum image.
%the image of the continuum {\bf emission}. 
The star is associated with a marginally resolved emission component with a 0\farcs2$\times$0\farcs15 beam. We use the IMFIT routine within CASA to fit two-dimensional Gaussians to the continuum data and derive both continuum fluxes and disk sizes (deconvolved from the beam) for the resolved sources.
As a result, we derive a flux of 9.0$\pm${0.9} mJy and a FWHM for the emission region of (0\farcs067$\pm$0\farcs008)$\times$(0\farcs058$\pm$0\farcs008) in the image plane. The latter corresponds to 100$_{-24}^{+35}$$\times$87$_{-22}^{+32}$ au including the uncertainty in the distance.

The centroid position measured using Gaussian fitting is ($\alpha_{2000}$,$\delta_{2000}$) = (06:57:22.221,$-$8:23:17.67), coinciding with the position of the 2MASS point source within the accuracy of the 2MASS measurement ($\pm$0\farcs06). We do not find any other clear emission components within 19\arcsec~of the continuum source. The 3-$\sigma$ detection limits for the point source are 0.15 mJy at the source, and 0.34 mJy at 15\arcsec~away from the source.

  %%%%%%%%%%%%%%%%%%%%%%%%%%%%%%%%%%%%%%%%
  %% Figure: continuum
  %%%%%%%%%%%%%%%%%%%%%%%%%%%%%%%%%%%%%%%%
  
  %%%%%%%%%%%%%%%%%%%%%%%%%%%%%%%%%%%%%%%%
  %%%%%%%%%%%%%%%%%%%%%%%%%%%%%%%%%%%%%%%%
  %% CO
  %%%%%%%%%%%%%%%%%%%%%%%%%%%%%%%%%%%%%%%%
  %%%%%%%%%%%%%%%%%%%%%%%%%%%%%%%%%%%%%%%%

\subsection{CO Emission in a wide FOV (13\arcsec$\times$13\arcsec)} \label{sec:results:vcm1}

\begin{figure}[ht!]
\epsscale{1.2}
\plotone{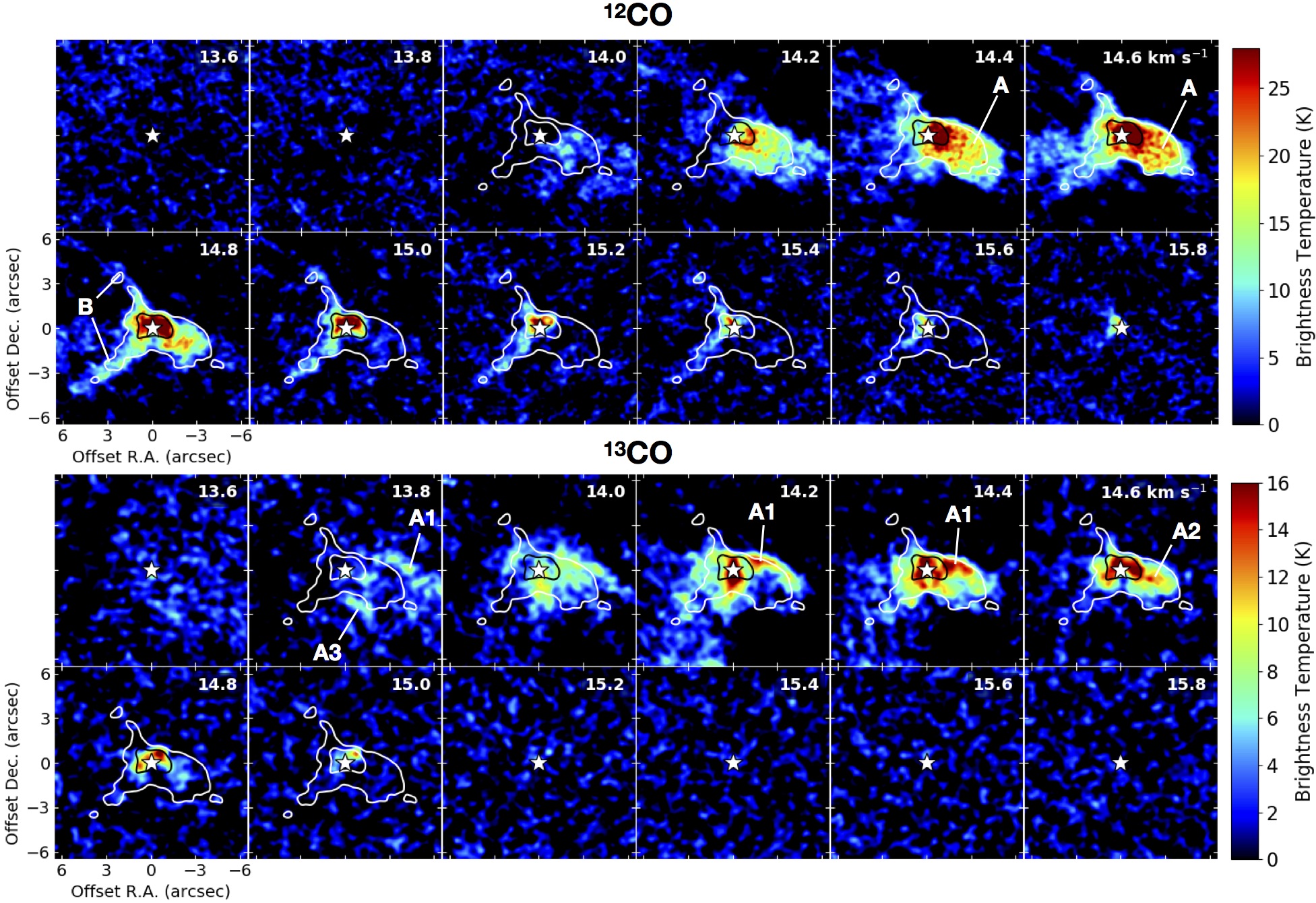}
\caption{The velocity channel maps for the $^{12}$CO and $^{13}$CO lines, with a velocity resolution of 0.2 km s$^{-1}$. The velocity at the top-right of each map is shown in the local standard rest frame. The (0\arcsec,0\arcsec) position, which is also shown in each map using the star symbol, is the centroid position of the continuum (Section \ref{sec:results:continuum}). A few arbitrary contours of the $^{12}$CO moment 0 map {(7.7 and 23.0 K km s$^{-1}$)} are shown to clarify the different spatial distributions between different lines and velocities. {The color of each contour (white/black) is arbitrarily chosen to highlight it against the color map.}
The features *A*, *A1*, *A2*, *A3* and *B* discussed in the text are also marked. 
\label{fig:vcm:1213CO}}
\end{figure}

\begin{figure}[ht!]
\epsscale{1.2}
\plotone{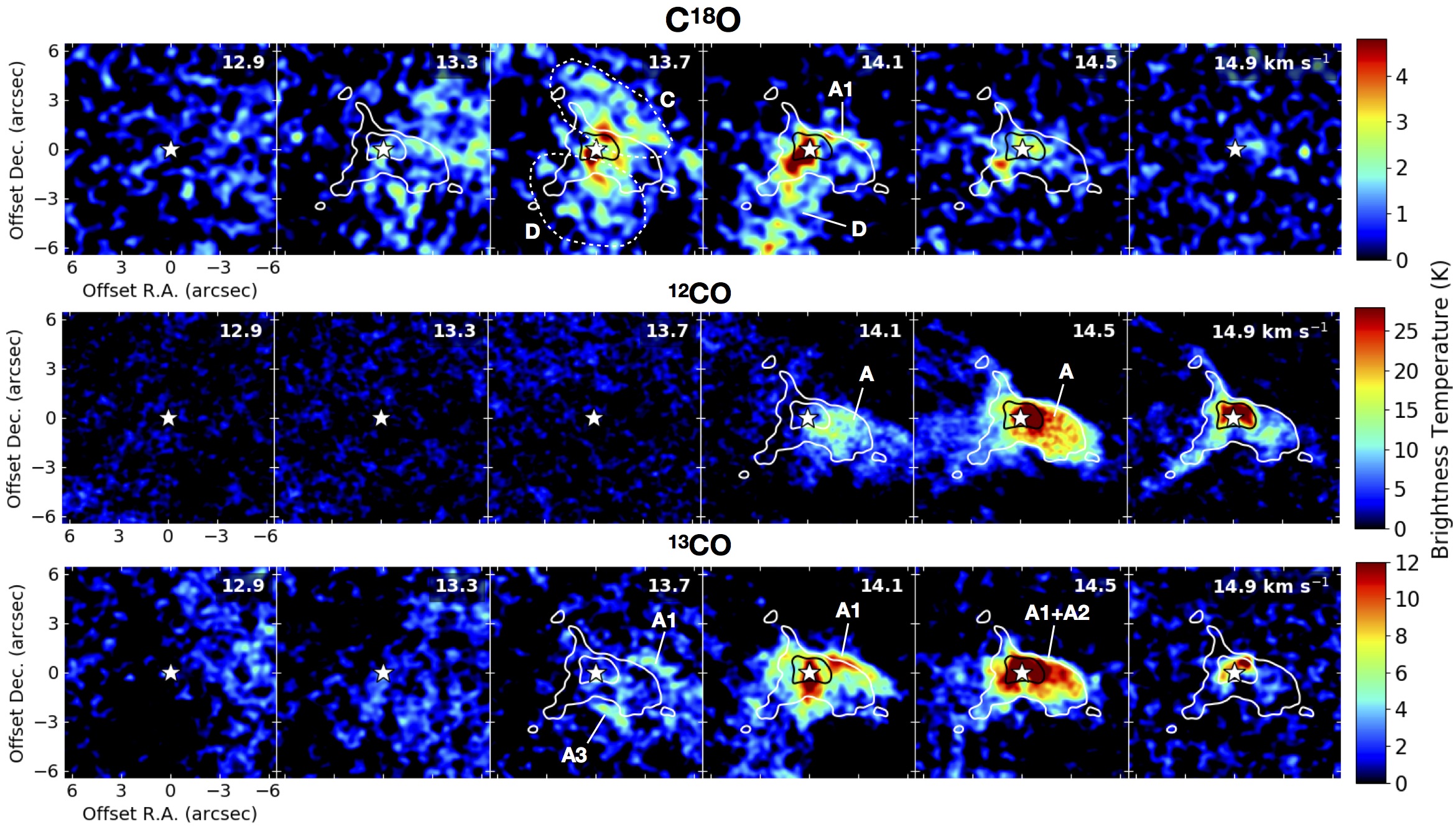}
\caption{
Same as Figure \ref{fig:vcm:1213CO} but for the C$^{18}$O (top), $^{12}$CO (middle), and $^{13}$CO lines (bottom) at 12.9--14.9 km s$^{-1}$ (i.e., those we identify the C$^{18}$O emission) with a velocity resolution of 0.4 km s$^{-1}$. The features discussed in the text are also marked. 
\label{fig:vcm:C18O}}
\end{figure}

\begin{figure}[ht!]
\epsscale{1.2}
\plotone{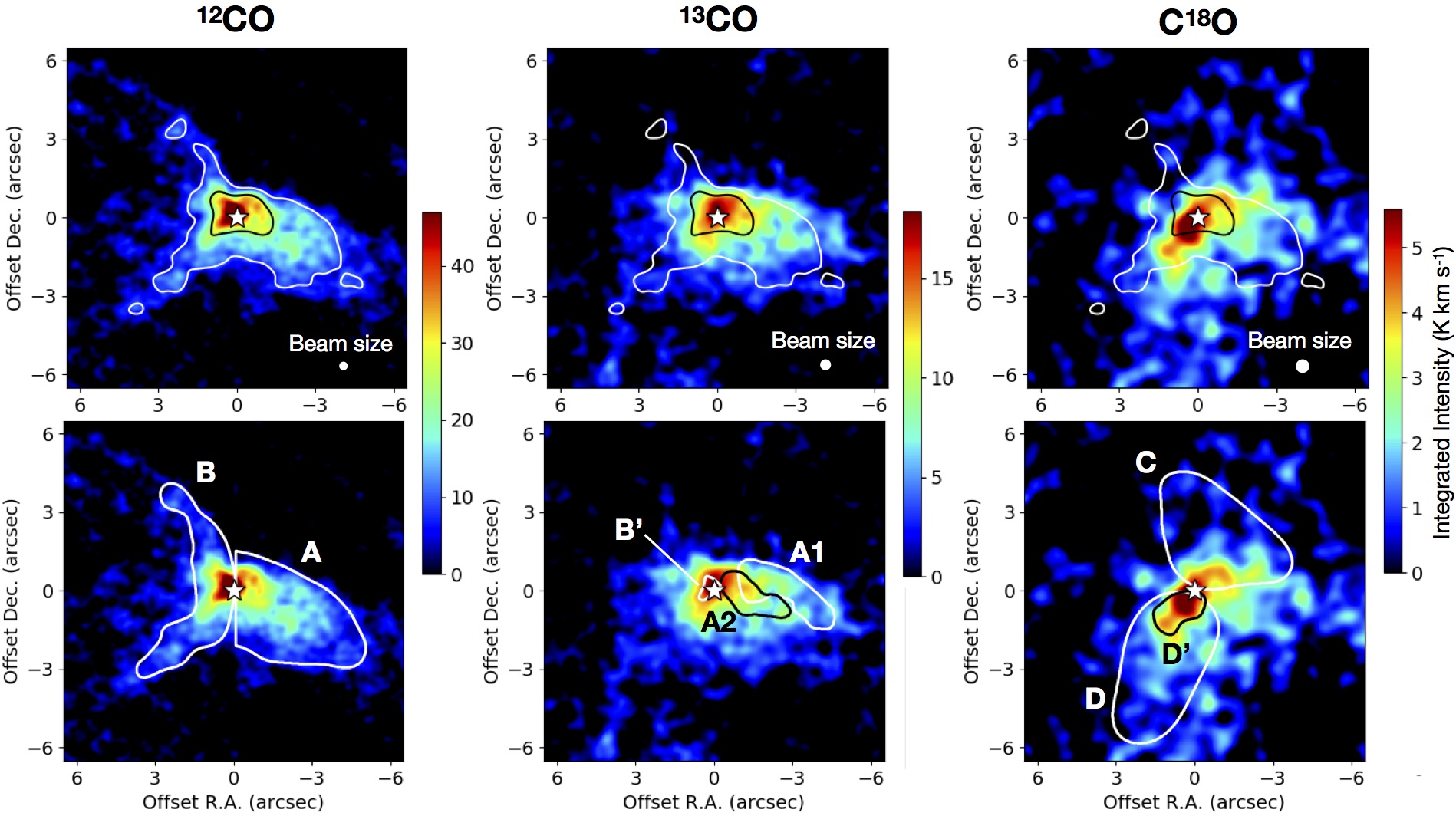}
\caption{The moment 0 maps of the $^{12}$CO, $^{13}$CO, and C$^{18}$O lines. The color scale between the top and bottom panels is identical. In the top panels we show the contours of the $^{12}$CO emission as for Figures \ref{fig:vcm:1213CO} and \ref{fig:vcm:C18O} to clarify the different spatial distributions between different lines. The bottom panels show the areas of the individual features (see the text and Figures \ref{fig:vcm:1213CO}, \ref{fig:vcm:C18O}, \ref{fig:vcm:1213CO:zoom-in}, and \ref{fig:vcm:C18O:zoom-in}) used for extracting the line profiles.
\label{fig:mom0}}
\end{figure}

Figures \ref{fig:vcm:1213CO} and \ref{fig:vcm:C18O} show the velocity channel maps of the $^{12}$CO, $^{13}$CO, and C$^{18}$O lines in a 13\arcsec$\times$13\arcsec~FOV (i.e., up to $\sim$10000 au scale from the continuum source). In Figure \ref{fig:vcm:1213CO} we show the maps for the $^{12}$CO and $^{13}$CO lines, which have relatively high signal-to-noise, with a 0.2 km s$^{-1}$ resolution. The signal-to-noise of the C$^{18}$O line is significantly lower, therefore we show the velocity channel maps of this line in Figure \ref{fig:vcm:C18O} with a 0.4 km s$^{-1}$ resolution. In Figure \ref{fig:vcm:C18O} we also show the maps for the $^{12}$CO and $^{13}$CO lines at the same velocities for comparison. Figure  \ref{fig:mom0} shows the moment 0 maps of the three lines in the same field of view. For each figure we show a few contours based on the moment 0 map of $^{12}$CO to clarify the different emission distributions between different lines and velocities.

  %%%%%%%%%%%%%%%%%%%%%%%%%%%%%%%%%%%%%%%%
  %% Figure: Velocity channel maps (12CO, 13CO)
  %%%%%%%%%%%%%%%%%%%%%%%%%%%%%%%%%%%%%%%%

  %%%%%%%%%%%%%%%%%%%%%%%%%%%%%%%%%%%%%%%%
  %% Figure: Velocity channel maps (C18O vs. 12CO, 13CO)
  %%%%%%%%%%%%%%%%%%%%%%%%%%%%%%%%%%%%%%%%

  %%%%%%%%%%%%%%%%%%%%%%%%%%%%%%%%%%%%%%%%
  %% Figure: Moment 0 maps
  %%%%%%%%%%%%%%%%%%%%%%%%%%%%%%%%%%%%%%%%

The $^{12}$CO maps show the following prominent features: *A* is an oval feature that extends from the continuum position to the west at a $\sim$5\arcsec~($\sim$8000 au) scale at $v_{\mathrm LSR}$=14--15 km s$^{-1}$; and *B* an arc to the east of the continuum position at $v_{\mathrm LSR}$=14.2--15.6 km s$^{-1}$ at a 5\arcsec--7\arcsec~ ($\sim$7000-11000 au) scale. As discussed later in Section 4.1 in detail, these are probably associated with blueshifted and redshifted molecular outflows, respectively.

Figure \ref{fig:vcm:1213CO} shows that the blueshifted outflow *A* is also associated with $^{13}$CO emission, but observed only at the northern edge at 13.8--14.4 km s$^{-1}$ (*A1*), the axis of the outflow at 14.6 km s$^{-1}$ (*A2*), and the southern edge at 13.8 km s$^{-1}$ (*A3*). These features are also seen in Figure \ref{fig:vcm:C18O} in the $^{13}$CO emission at 13.7, 14.1 and 14.5 km s$^{-1}$, and in the C$^{18}$O emission at 14.1 km s$^{-1}$. We see no $^{13}$CO emission associated with the feature *B*.

In Figure \ref{fig:vcm:C18O} the C$^{18}$O emission shows diffuse extended emission in the north-south direction at 13.3, 13.7, and 14.1 km s$^{-1}$ at a $\sim$5 arcsec ($\sim$8000 au) scale. The northern component (*C*) is seen at 13.3 and 13.7 km s$^{-1}$; while the southern component (*D*) is seen at 13.7 and 14.1 km s$^{-1}$. At 14.1 km s$^{-1}$ the component *D* extends to the outside of the FOV of the figure, up to $\sim$10 arcsec from the star.
The C$^{18}$O emission does not show *A*, *A1*, *A2*, *A3*, or *B* seen in the $^{12}$CO and $^{13}$CO emission except the feature *A1* at 14.1 km s$^{-1}$.

  %%%%%%%%%%%%%%%%%%%%%%%%%%%%%%%%%%%%%%%%
  %% Sub-Section : CO, bright compact components
  %%%%%%%%%%%%%%%%%%%%%%%%%%%%%%%%%%%%%%%%

\subsection{CO Emission Close to the Continuum Source (4\arcsec$\times$4\arcsec~FOV)} \label{sec:results:vcm2}

\begin{figure}[ht!]
\epsscale{1.2}
\plotone{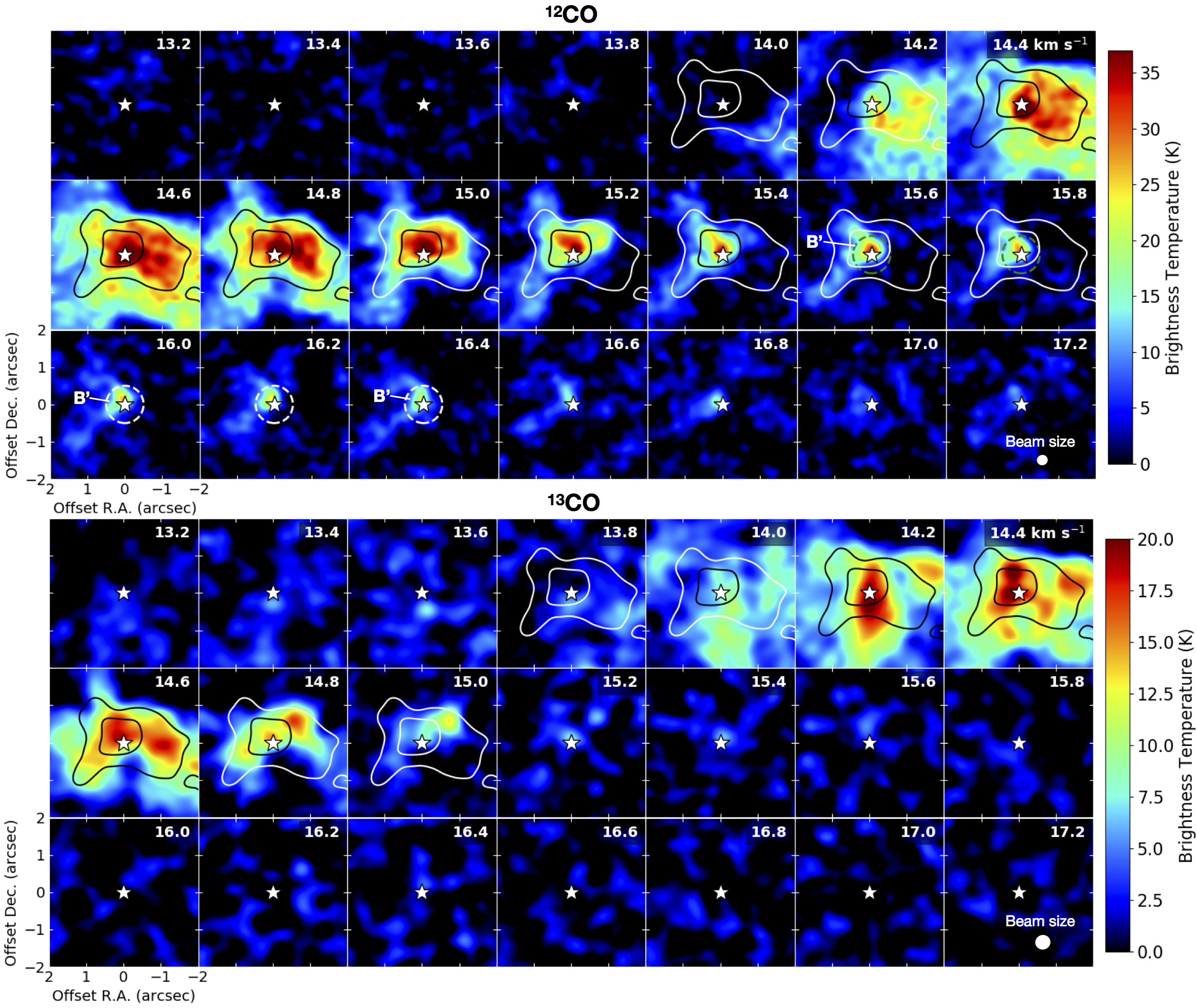}
\caption{Same as Figure \ref{fig:vcm:1213CO} but in a 4\arcsec$\times$4\arcsec~FOV. {The contour levels are 19 and 35 K km s$^{-1}$ in the $^{12}$CO emission.} In some $^{12}$CO maps we draw a dashed circle centered on the continuum source, 0\farcs5 in radius, to clarify the angular scale of the compact feature *B'*. 
\label{fig:vcm:1213CO:zoom-in}}
\end{figure}

\begin{figure}[ht!]
\epsscale{1.2}
\plotone{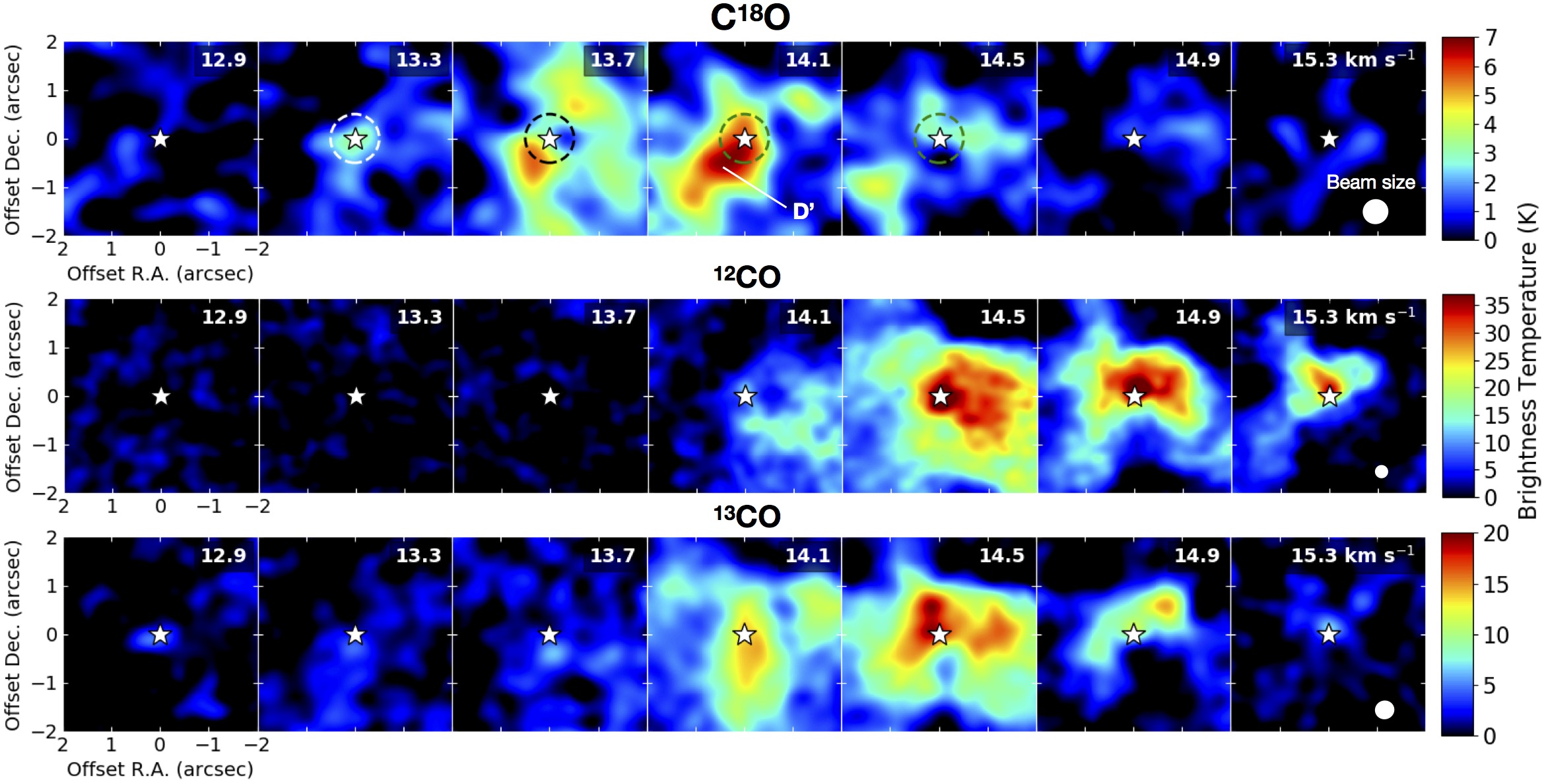}
\caption{
Same as Figure \ref{fig:vcm:C18O} but in a 4\arcsec$\times$4\arcsec~FOV, and without contours for the $^{12}$CO moment 0 map. In some $^{18}$CO maps we draw a dashed circle centered on the continuum source, 0\farcs5 in radius, for reference.
\label{fig:vcm:C18O:zoom-in}}
\end{figure}

In Figures \ref{fig:vcm:1213CO}--\ref{fig:mom0}, all the lines show bright compact emission close to the continuum position. Figures \ref{fig:vcm:1213CO:zoom-in} and \ref{fig:vcm:C18O:zoom-in} show the velocity channel maps for a 4\arcsec$\times$4\arcsec~FOV: i.e., up to $\sim$3000 au scale from the continuum source. The figures show the $^{12}$CO emission at $v_\mathrm{LSR}$=14.2--16.8 km s$^{-1}$, the $^{13}$CO emission at 14.0--15.0 km s$^{-1}$, and the C$^{18}$O emission at 13.3--14.5 km s$^{-1}$.
Bright $^{12}$CO emission at 14.2--14.8 km s$^{-1}$ is explained as the base of the blueshifted outflow *A*. The $^{12}$CO emission is fainter at larger velocities, and its peak is offset to the north by up to $\sim$0\farcs5 at 14.8--15.2 km s$^{-1}$. The maps for 15.4--16.4 km s$^{-1}$ show a compact component at the base of the arc *B*,  whose peak is marginally offset to the {northeast} by $\sim$0\farcs1 (hereafter *B'*).

  %%%%%%%%%%%%%%%%%%%%%%%%%%%%%%%%%%%%%%%%
  %% Figure: Velocity channel maps (12CO, 13CO), zoom-in
  %%%%%%%%%%%%%%%%%%%%%%%%%%%%%%%%%%%%%%%%

  %%%%%%%%%%%%%%%%%%%%%%%%%%%%%%%%%%%%%%%%
  %% Figure: Velocity channel maps (C18O vs. 12CO, 13CO), zoom-in
  %%%%%%%%%%%%%%%%%%%%%%%%%%%%%%%%%%%%%%%%

In Figures \ref{fig:vcm:1213CO:zoom-in} and \ref{fig:vcm:C18O:zoom-in} the $^{13}$CO and C$^{18}$O emission show kinematic structures significantly different from the $^{12}$CO emission. In particular, the $^{13}$CO emission in these figures  show complicated intensity distributions at different velocities. The maps for 14.6 km s$^{-1}$ in Figure \ref{fig:vcm:1213CO:zoom-in} and 14.5 km s$^{-1}$ in Figure \ref{fig:vcm:C18O:zoom-in}, show bright extended emission to the west, which may correspond to the base of the feature *A2* shown in Figures \ref{fig:vcm:1213CO} and \ref{fig:vcm:C18O}. The maps for 14.2 km s$^{-1}$ in Figure \ref{fig:vcm:1213CO:zoom-in} and 14.1 km s$^{-1}$ in Figure \ref{fig:vcm:C18O:zoom-in} show bright extended emission in the north-south direction. The C$^{18}$O line at 13.7 and 14.1 km s$^{-1}$ also shows extended emission in the north-south direction, which may be the base of the emission components *C* and *D* in Figure \ref{fig:vcm:C18O}. The brightest component (*D'*) is seen at 14.1 km s$^{-1}$ in the southeast direction with a spatial scale up to 2--3\arcsec, peaking at $\sim$0\farcs5 from the continuum position.

  %%%%%%%%%%%%%%%%%%%%%%%%%%%%%%%%%%%%%%%%
  %% Sub-Section : Line profiles
  %%%%%%%%%%%%%%%%%%%%%%%%%%%%%%%%%%%%%%%%

\subsection{CO Line Profiles} \label{sec:results:profiles}

\begin{figure}[ht!]
\epsscale{1}
\plotone{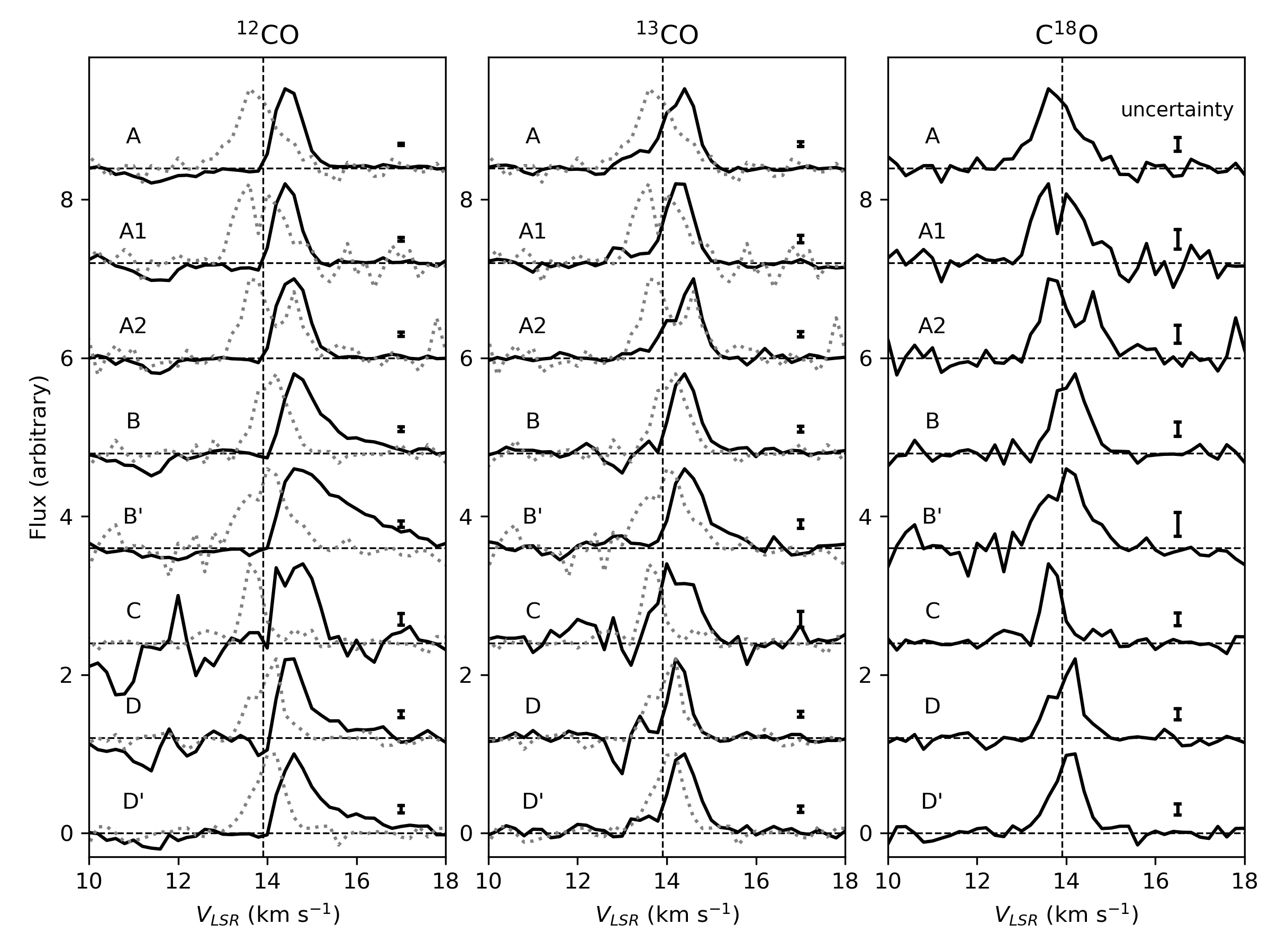}
\caption{The $^{12}$CO, $^{13}$CO, and C$^{18}$O line profiles extracted for the individual features discussed in the text. Each line profile is normalized to the peak, and arbitrarily offset.
The uncertainty of each profile is shown on the right. For $^{12}$CO and $^{13}$CO, we overplot the C$^{18}$O profiles at the same position for comparison with gray dotted curves. The vertical dashed line shows the reference velocity (13.9 km s$^{-1}$, see text).
\label{fig:line-profs:all}}
\end{figure}

Figure \ref{fig:line-profs:all} shows the profiles of the three CO lines at the features discussed in Sections \ref{sec:results:vcm1} and \ref{sec:results:vcm2}. The areas used for extracting the line profiles are shown in the bottom panels of Figure \ref{fig:mom0}. We exclude *A3* due to low signal-to-noise.

  %%%%%%%%%%%%%%%%%%%%%%%%%%%%%%%%%%%%%%%%
  %% Figure: Line profiles (all)
  %%%%%%%%%%%%%%%%%%%%%%%%%%%%%%%%%%%%%%%%

The $^{12}$CO emission shows a relatively sharp cutoff of the flux at $V_\mathrm{LSR}$$\sim$14 km s$^{-1}$, and we observe zero or negative fluxes at smaller velocities. We attribute this to a combination of the following: (1) a contribution from extended emission associated with the parent cloud, which does not allow us to clearly observe emission associated with the YSO with large spatial frequencies; and (2) modest $uv$ coverages of the observations. The $^{13}$CO emission shows the same trend to some extent, but this line still shows weak fluxes at $V_\mathrm{LSR}$$<$14 km s$^{-1}$. In contrast, the C$^{18}$O emission does not show this trend. Such differences are explained by different abundance of C and O isotopes \citep[$\mathrm{^{12}C}$/$\mathrm{^{13}C}$$\sim$70 and $\mathrm{^{16}O}$/$\mathrm{^{18}O}$$\sim$490 for the local interstellar medium;][]{Mathis00}, and therefore different molecular abundances. In other words, a low abundance of C$^{18}$O (and $^{13}$CO) make the extended emission associated with the molecular cloud significantly fainter, allowing the emission from the YSOs to be observed at these velocities. At $V_\mathrm{LSR}$$>$14 km s$^{-1}$, the $^{12}$CO line profiles shown at *B*, *B'*, *D*, and *D'* are associated with an excessive redshifted wing up to $V_\mathrm{LSR}$$\sim$19 km s$^{-1}$

%We plot the C$^{18}$O profiles in Figure \ref{fig:line-profs:C18O} for all the features in the same panel.
We tabulate the peak and FWHM velocities for the C$^{18}$O emission in Table \ref{tab:prof_params}. We do not know the systemic velocity of the target exactly, therefore we describe the blue- and red-shift of the emission using ``the reference velocity'', which we tentatively define as 13.9 km s$^{-1}$. The peaks associated with the features *A* and *B* are blueshifted and redshifted from the reference velocity by $\sim$0.3 km s$^{-1}$. The features *A1* and *A2*, which are observed in *A*, show two peaks at the blueshifted and redshifted sides, respectively, with $\Delta v$=0.1-0.7 km s$^{-1}$ from the reference velocity. The blueshifted emission seems brighter than the redshifted emission for both profiles. The profiles for *B'*, *C*, *D*, and *D'* show a single peak, offset from the reference velocity by 0.1-0.3, --(0.1-0.3), $\sim$0.3 and 0.1-0.3 km s$^{-1}$, respectively. The FWHM velocities measured at *A*, *A1*, *A2*, *B* are 0.8--1.3 km s$^{-1}$, larger than those for *C*, *D*, and *D'* (0.5--0.8 km s$^{-1}$).

  %%%%%%%%%%%%%%%%%%%%%%%%%%%%%%%%%%%%%%%%
  %% Figure: Line profiles (C18O)
  %%%%%%%%%%%%%%%%%%%%%%%%%%%%%%%%%%%%%%%%

%\begin{figure*}
%\epsscale{0.35}
%\plotone{f_C18O_profs.png}
%\caption{The C$^{18}$O line profiles extracted for the individual features discussed in the text. Each line profile is normalized to the peak, and arbitrarily offset. The dashed line shows the reference velocity (13.9 km s$^{-1}$, see text).
%\label{fig:line-profs:C18O}}
%\end{figure*}

  %%%%%%%%%%%%%%%%%%%%%%%%%%%%%%%%%%%%%%%%
  %% Table : Convolution
  %%%%%%%%%%%%%%%%%%%%%%%%%%%%%%%%%%%%%%%%
  
\begin{deluxetable*}{ccc}[b!]
\tablecaption{Peak and FWHM Velocities for C$^{18}$O Emission \label{tab:prof_params}}
\tablecolumns{3}
%\tablenum{2}
\tablewidth{0pt}
\tablehead{
\colhead{Feature/Area\tablenotemark{a}} &
\colhead{$V_\mathrm{Peak}$ (km s$^{-1}$)} &
\colhead{$V_\mathrm{FWHM}$ (km s$^{-1}$)} 
}
\startdata
A	& 13.6			& 0.9 \\
A1	& 13.6/14.0		& 1.2 \\
A2	& 13.6-13.8/14.6 	& 1.3 \\
B	& 14.2			& 0.8 \\
B'	& 14.0-14.2		& 1.1 \\
C	& 13.6-13.8		& 0.5 \\
D	& 14.2			& 0.8 \\
D'	& 14.0-14.2		& 0.8\vspace{0.2cm}\\
N2	& 13.6			& 0.2 \\
N1	& 13.6-13.8		& 0.6 \\
S1	& 14.0-14.2		& 0.9 \\
S2	& 14.2			& 0.3 
\enddata
\tablenotetext{a}{See Sections 3 and 4.1 for the definition of A-D' and N1/N2/S1/S2, respectively.}
\end{deluxetable*}

%%%%%%%%%%%%%%%%%%%%%%%%%%%%%%%%%%%%%%%%
%%%%%%%%%%%%%%%%%%%%%%%%%%%%%%%%%%%%%%%%
%%%%%%%%%%%%%%%%%%%%%%%%%%%%%%%%%%%%%%%%
%% Section : Discussion (1)
%%%%%%%%%%%%%%%%%%%%%%%%%%%%%%%%%%%%%%%%
%%%%%%%%%%%%%%%%%%%%%%%%%%%%%%%%%%%%%%%%
%%%%%%%%%%%%%%%%%%%%%%%%%%%%%%%%%%%%%%%%

\section{Extended molecular outflows, Envelope, Disk and Hot Wind\footnote{Please refer the footnote in Section 1 for use of the words ``outflow'' and ``wind".}} \label{sec:discussion}

  %%%%%%%%%%%%%%%%%%%%%%%%%%%%%%%%%%%%%%%%
  %%%%%%%%%%%%%%%%%%%%%%%%%%%%%%%%%%%%%%%%
  %% Subsection : Bipolar Outflow
  %%%%%%%%%%%%%%%%%%%%%%%%%%%%%%%%%%%%%%%%
  %%%%%%%%%%%%%%%%%%%%%%%%%%%%%%%%%%%%%%%%

In Sections \ref{sec:discussion:outflow} and \ref{sec:discussion:envelope} we discuss CO emission associated with the extended molecular outflows and the envelope, respectively. In Section \ref{sec:discussion:wind} we discuss the possible interaction of the hot FUor wind seen in optical spectra with the circumstellar environment. In Section \ref{sec:discussion:continuum} we discuss the compact nature of the continuum emission, which is primary attributed to the circumstellar disk but with a possible contribution from an FUor wind or a wind cavity.

\subsection{Extended Molecular Bipolar outflow} \label{sec:discussion:outflow}

The $^{12}$CO, $^{13}$CO and C$^{18}$O channel maps show remarkably different intensity distributions even at the same velocities. Again, this is explained by different optical thicknesses of these lines due to different molecular abundances. The $^{12}$CO emission can be optically thick in general, and as a result, the observed fluxes highly depend on temperature distributions. In contrast, the optically thin C$^{18}$O emission is significantly affected by the column density distribution as well as the temperature distribution.

Optical and near-IR imaging observations by \citet{Reipurth12} showed the presence of a reflection nebula extending toward the southwest of the star at 20--50 arcsec. In their near-IR image the nebula extends to the west within 2-3 arcsec of the star, i.e., the same direction as *A* in $^{12}$CO emission shown in Section \ref{sec:results:vcm1}. Reflection nebulae with similar morphologies are often associated with an outflow cavity in the protostellar envelope associated with many YSOs \citep[e.g.,][]{Tamura91,Lucas96,Lucas98a,Padgett99}. The outflows associated with YSOs are usually bipolar \citep[see, e.g.,][for a review]{Arce07}, and the absence of the counterflow component at the optical and near-IR reflection nebulae is explained by a larger extinction towards the redshifted outflow.

Therefore, we conclude that the features *A* and *B* in Figures \ref{fig:vcm:1213CO}-\ref{fig:mom0} are associated with blueshifted and redshifted lobes of molecular outflows, respectively. The feature *A* does not show blueshifted $^{12}$CO emission, presumably due to the parent molecular cloud (Section \ref{sec:results:profiles}). Even so, the C$^{18}$O line profiles for *A* and *B* show blue-and red-shifts (Figure \ref{fig:line-profs:all} and Table \ref{tab:prof_params}), agreeing with this interpretation. The presence of the redshifted $^{12}$CO emission in the blueshifted outflow lobe is explained: (1) by the fact that the millimeter CO emission in molecular outflows results from interaction of a collimated jet or a wide-angled wind with the surrounding gas \citep[e.g.,][]{Lee02,Arce07}, and (2) if the $^{12}$CO emission in feature *A* is associated with the far side of the expanding outflow lobe.

The $^{13}$CO line shows emission at the northern and southern edges of the blueshifted lobe of the outflow *A* (*A1* and *A3* in Figures \ref{fig:vcm:1213CO} and \ref{fig:vcm:C18O}), suggesting the presence of gas compressed at the outflow cavity. The feature *A2* in the $^{13}$CO emission could be attributed to gas compression at the far side of the outflow cavity.
The $^{13}$CO emission at the northern edge of the blueshifted outflow (*A1*) is significantly brighter than the southern edge (*A3*), suggesting that the jet/wind interaction with the surrounding gas is not symmetric. The enhanced emission in the northern edge may be explained if the orientation of the jet/wind that drives the extended molecular outflow gradually moves counterclockwise: i.e., southwest in the past, and west at present. This speculation is consistent with the fact that the reflection nebula in the outer region ($>$10\arcsec~of the star) is located in the southwest. Interestingly, we do not find such a brightness asymmetry between the northern and the southern sides of the counter outflow *B*.

In contrast to some extended CO outflows, which show a fairly good symmetry between the blueshifted and redshifted outflow lobes with respect to the YSO \citep[e.g.,][]{Arce06}, the blueshifted outflow lobe *A* is significantly brighter than the redshifted outflow lobe. This may be due to different densities of the ambient gas between the blueshifted and redshifted sides, as with a blue-/red-ward asymmetry observed in some CO outflows \citep{Lee02}. The single-dish observations of the molecular core by \citet{Kospal17b} show that the center of the core (and therefore the densest region) is located at {$\sim$30\arcsec~west} from the target YSO. This fact would also explain the different opening angles of the blueshited and redshifted lobes of the CO outflows, as more diffuse ambient gas requires less momentum to widen an outflow cavity, thereby making the opening angle of the redshifted lobe *B* wider. 
%However, this discussion would require emission at the entire velocity range, which is impossible for our data due to contaminating emission from the parent molecular cloud (Section \ref{sec:results:profiles}).

  %%%%%%%%%%%%%%%%%%%%%%%%%%%%%%%%%%%%%%%%
  %%%%%%%%%%%%%%%%%%%%%%%%%%%%%%%%%%%%%%%%
  %% Subsection : Envelope
  %%%%%%%%%%%%%%%%%%%%%%%%%%%%%%%%%%%%%%%%
  %%%%%%%%%%%%%%%%%%%%%%%%%%%%%%%%%%%%%%%%

\subsection{Molecular Envelope} \label{sec:discussion:envelope}

The diffuse C$^{18}$O emission components *C*, *D* and *D'* are presumably associated with a circumstellar envelope for the following reasons. These components are extended in the north-south direction, perpendicular to the  outflows *A* and *B* (Figures \ref{fig:vcm:1213CO} and \ref{fig:mom0}). The absence of these components in the $^{12}$CO and $^{13}$CO indicates their relatively high column densities and low temperatures. Furthermore, these emission components also have FWHM velocities smaller than the outflow components (Section \ref{sec:results:profiles}), exhibiting their relatively quiescent nature. All of these trends are what we expect for a circumstellar envelope associated with low-mass YSOs \citep[e.g.,][]{Stahler05}.

\begin{figure*}
\epsscale{1.0}
\plotone{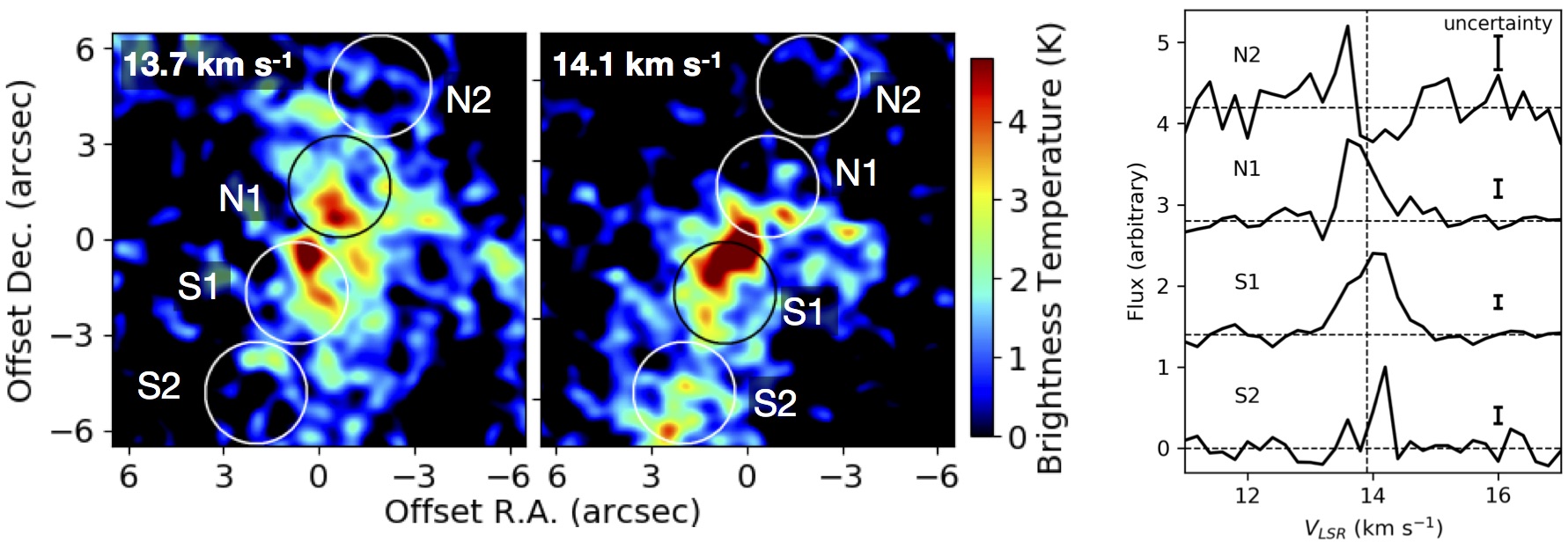}
\caption{(left) Areas for extracting line profiles, plotting over the velocity channel maps of the C$^{18}$O emission. (right) Line profiles. 
\label{fig:line-profs:C18O:envelope}}
\end{figure*}

The northern and southern sides of the envelope (i.e., *C* and *D*, respectively) are blue- and red-shifted, respectively (Section \ref{sec:results:profiles}, Figure \ref{fig:line-profs:all}), indicative of its rotational motion. To further investigate this motion, we derive the line profiles at four apertures along the envelope emission, 1\farcs8 and 5\farcs2 from the star, and plot them in Figure \ref{fig:line-profs:C18O:envelope}. Their peak and FWHM velocities are tabulated in Table \ref{tab:prof_params}. The peak velocity at outer radii (*N2* and *S2*) is $\sim$ 0.3 km s$^{-1}$ offset from the reference velocity. This is comparable to that expected for a Keplerian motion at the given distance ($\sim$8000 au at $d$=1.5 kpc) and assuming 1 $M_\sun$. In contrast, the peak offset velocity in the inner radii (0.1--0.3 km s$^{-1}$) is significantly lower than the Keplerian velocity ($\sim$0.6 km s$^{-1}$) expected at the given distance ($\sim$3000 au) and the same stellar mass. Therefore, the envelope gas up to 3000--5000 au could infall toward the star (see Section \ref{sec:uni:seds} for further discussion).

  %%%%%%%%%%%%%%%%%%%%%%%%%%%%%%%%%%%%%%%%
  %% Figure: Line profiles (C18O), envelope
  %%%%%%%%%%%%%%%%%%%%%%%%%%%%%%%%%%%%%%%%

%The deviation of the measured velocity field from the Keplerian motion may be due to {\bf weak interaction with the ambient magnetic field} in the inner region \citep[e.g.,][for a review]{Li14}. We note that, however, we should take into account the time scale required for stabilizing the kinematics \citep[2$\pi$$r$/$v$, a dynamical time scale;][]{Frank02}. Provided $\Delta v$$\sim$0.3 km s$^{-1}$, one would derive dynamical time scales of $\sim$0.3 and $\sim$0.8 Myrs at the above radii ($\sim$3000 and $\sim$8000 au, respectively). The latter may be too large for the age expected for the evolutionary stage of this YSO ($<$4-5$\times$10$^5$ yrs; Section \ref{sec:uni}). Therefore, the kinematics at the outer radii of the envelope may not have reached dynamical equilibrium. 

The southern part of the envelope (*D*, *D'*) is brighter than the northern part in Figures \ref{fig:vcm:C18O} and \ref{fig:vcm:C18O:zoom-in}, i.e., at both large (up to $\sim$10$^4$ au) and small spatial scales ($\lesssim$3000 au). 
A similar asymmetry at the $\gg$1000-au scales of the envelope CO emission has previously been observed for a few other FUors and FUor-like stars by \citet{Kospal17b,Feher17,Ruiz17_HBC494}.
The C$^{18}$O emission in Figure \ref{fig:vcm:C18O:zoom-in} shows peaks at the southeast and the northwest at 13.7 and 14.1 km s$^{-1}$, offset by $\sim$0\farcs5 ($\sim$800 au), without exhibiting an emission peak at the continuum position. This contrasts to the envelopes of Class 0-I YSOs, whose envelope emission peaks at the star \citep[e.g.,][]{Arce06}. One would estimate a dynamical scale of these components of $\sim$1$\times$10$^5$ yr with the given distance from the continuum source and the offset velocity ($\sim$0.2 km s$^{-1}$). Therefore, such an asymmetry in the innermost envelope would disappear within a fairly short time scale compared with the entire timescale of the envelope dissipation, i.e., the termination of the Class I phase \citep[4-5$\times$10$^5$ years, see][for a review]{Dunham14}. This fact suggests that this asymmetry in the innermost region is associated with an activity or phenomenon significantly shorter than the above dynamical timescale (see Section \ref{sec:discussion:wind}).

  %%%%%%%%%%%%%%%%%%%%%%%%%%%%%%%%%%%%%%%%
  %%%%%%%%%%%%%%%%%%%%%%%%%%%%%%%%%%%%%%%%
  %% Subsection :  Interaction with a powerful wind
  %%%%%%%%%%%%%%%%%%%%%%%%%%%%%%%%%%%%%%%%
  %%%%%%%%%%%%%%%%%%%%%%%%%%%%%%%%%%%%%%%%

\subsection{Interaction with a Hot FUor Wind?} \label{sec:discussion:wind}

\citet{Reipurth12} observed P Cygni profiles in optical permitted lines (H$\alpha$, Na D, Ca II) towards the star. These line profiles indicate the presence of a hot ($\gg$1000 K) energetic wind ($v \sim 200$ km s$^{-1}$) as for many other FUors \citep{Hartmann96,Audard14}. How is this FUor wind interacting with the extended molecular outflows and the envelope seen in the CO emission?

As reported by \citet{Thommes11,Reipurth12}, the outburst began between 1953 and 2009. 
%Therefore, the {\bf hot FUor} wind {\bf from the star or an inner disk} would reach a distance of up to $\sim$2700 au ($\sim$2 arcsec) from the star in 65 years, with the given velocity of $v \sim 200$ km s$^{-1}$ and assuming that the {\bf hot} wind seen in the optical line profiles emerged at the onset of the outburst.  However, the intensity distributions in the {\bf extended} CO outflows and the envelope are continuous over this spatial scale (see Section \ref{sec:results:vcm1}), without any clear evidence for interaction with the {\bf hot} energetic FUor wind.
If we assume that the wind seen in the optical line profiles emerged at the onset of the outburst, we would expect that in 65 years the wind would have travelled $\sim$2700 au ($\sim$2 arcsec). However, none of the CO lines exhibit clear evidence of the interaction between the wind and the surrounding material within this 2 arcsec window.
%This implies that we can discuss the nature of the progenitor of the outbursts with the observed  CO outflows and the envelope (Section \ref{sec:discussion2}).

In contrast, within $\sim$0\farcs5 ($\sim$800 au) of the continuum emission, the $^{12}$CO line at the base of the redshifted outflow (*B'*) shows the presence of bright high-velocity emission up to $\sim$19 km s$^{-1}$ (Figures \ref{fig:vcm:1213CO:zoom-in} and \ref{fig:line-profs:all}). Furthermore, the $^{18}$CO emission in Figure \ref{fig:vcm:C18O:zoom-in} is relatively faint within $\sim$0\farcs5 compared with the outer region. These may be due to interaction with the hot FUor wind:  i.e. high-velocity $^{12}$CO emission due to acceleration by the FUor wind, and a deficit of C$^{18}$O due to a cavity opened by the FUor wind.

In summary, interactions between the hot FUor wind and the circumstellar environment is limited to within 0\farcs5 (corresponding to $\sim$800 au) of the star at worst. The FUor wind has emerged relatively recently, considering the fact that a wind with $v \sim 200$ km s$^{-1}$ would require only $\sim$20 years to reach this distance. This fact suggests that the FUor outburst associated with this star has begun relatively recently, i.e., closer to 2009 than 1953 as constrained by the time range provided by optical imaging observations by \citet{Thommes11,Reipurth12}.

  %%%%%%%%%%%%%%%%%%%%%%%%%%%%%%%%%%%%%%%%
  %%%%%%%%%%%%%%%%%%%%%%%%%%%%%%%%%%%%%%%%
  %% Section : Dust continuum and disk
  %%%%%%%%%%%%%%%%%%%%%%%%%%%%%%%%%%%%%%%%
  %%%%%%%%%%%%%%%%%%%%%%%%%%%%%%%%%%%%%%%%

\subsection{Continuum and Circumstellar Disk} \label{sec:discussion:continuum}

We measured a size of the compact dust continuum emission at the star of 100$_{-24}^{+35}$$\times$87$_{-22}^{+32}$  au (Section \ref{sec:results:continuum}). With the measured flux of 9.0$\pm${0.9} mJy, we derive a typical temperature of the disk of 60--{120} K assuming an optically thick disk with a uniform temperature. The inferred temperature is higher if the filling factor of the area is less than the unity due to a low optical thickness, or a gravitational fragmentation of the disk (Section \ref{sec:intro}).
In Table \ref{tab:ALMA_FUors} we compare the observed flux and the angular scale for V900 Mon with ALMA observations of several other FUor and FUor-like stars at the same wavelength \citep{Kospal17c,Cieza18}. The table shows that these parameters for V900 Mon are similar to other FUors except for the large spatial scale observed for V346 Nor.

  %%%%%%%%%%%%%%%%%%%%%%%%%%%%%%%%%%%%%%%%
  %% Table : ALMA 230 GHz observations of FUors
  %%%%%%%%%%%%%%%%%%%%%%%%%%%%%%%%%%%%%%%%
  
\begin{deluxetable*}{cccccccccc}[b!]
\tablecaption{ALMA observations of the 1.3-mm continuum associated with FUor disks \label{tab:ALMA_FUors}}
%\tablecolumns{5}
%\tablenum{1}
\tablewidth{0pt}
\tablehead{
\colhead{Object} & 
\colhead{Category} &
\colhead{$L_{\mathrm bol}$} &
\colhead{$d$} &
\colhead{Onset} &
\multicolumn{2}{c}{$F_{\mathrm 1.3mm}$ (mJy)} &
\colhead{FWHM\tablenotemark{c}} &
\colhead{Resolution\tablenotemark{d}} &
%\colhead{Angular Scale}
\colhead{Reference}
\\
\colhead{} &
\colhead{} &
\colhead{($L_\sun$)} &
\colhead{(kpc)} &
\colhead{(yr)} &
\colhead{Measured} &
\colhead{$d$=1 kpc\tablenotemark{b}} &
\colhead{(au)} &
\colhead{(au)} 
}
\startdata
V900 Mon	& FUor		& $\sim$200\tablenotemark{a}	& 1.5	 & 1953--2010	& 9.0$\pm${0.9}	& {20$\pm$2}
	& $100_{-24}^{+35} \times 87_{-22}^{+32}$	& 600$\times$450	& This work \\
V346 Nor	& FUor		& 135	& 0.7		& $\sim$1980	& 27$\pm$3	& 13$\pm$1	
			& $420 \times 322$	& 770$\times$630	& \citet{Kospal17c}	\\
V883 Ori	& FUor-like	& 400	& 0.41	& ---			& 353$\pm$35	& 59$\pm$7	
			& $124_{-9}^{+9} \times 100_{-8}^{+8}$	& 103$\times$70	& \citet{Cieza18}	\\
HBC 494	& FUor-like	& 300	& 0.41	& ---			& 113$\pm$11	& 20$\pm$2	
			& $58_{-5}^{+5} \times 19_{-5}^{+5}$	& 103$\times$70	& \citet{Cieza18}	\\
V2775 Ori  & FUor/EXor	& $\sim$25	& 0.41	& 2005-2007	& 106$\pm$10	& 18$\pm$2	
			& $61_{-4}^{+4} \times 59_{-5}^{+5}$	& 103$\times$70	& \citet{Cieza18}	\\
\enddata
\tablenotetext{a}{We scale the luminosity measured by \citet{Reipurth12} to the recent measurement of the distance based on Gaia DR2 (Section \ref{sec:intro}).}
\tablenotetext{b}{Scaled to the given distance. The uncertainty of the distance is not included.}
\tablenotetext{c}{Measured by fitting the continuum emission and the beam using two-dimensional Gaussians in the image plane.}
\tablenotetext{d}{Spatial resolution of the observations. The uncertainty of the distance is not included.}
\end{deluxetable*}

Compact millimeter continuum emission associated with FUors is often attributed to a circumstellar disk \citep[e.g.,][]{Kospal17c,Cieza18,Liu18}. All the targets in Table 2 except V346 Nor would infer a disk radius below 100 au. To further investigate the nature of the millimeter emission, \citet{Cieza18} conducted radiative transfer modeling of passive disks, for which the heating source of the disk is radiation from the central star. These authors used a small radius for the disk (20--40 au) to reproduce their {millimeter} observations.

In contrast, \citet{Liu16, Takami18} conducted near-IR observations of scattered light in five well-studied FUors. Some of these objects show that scattered light, presumably associated with a disk surface, extends over 400-800 au, i.e., significantly larger than the radii of the millimeter continuum emission described above. Therefore, the compact nature of the millimeter disk emission would require the following physical processes:
%\\
%

{\it (A) Viscous disk accretion in the inner disk region} ---  This physical process, which is not included in the models by \citet{Cieza18}, heats up the FUor disk, in particular in the inner region, and enhance the radiation from the inner disk \citep[see][for a review]{Hartmann96}. As a result, the observed FWHM size of the entire emission region would be smaller. In Appendix \ref{sec:appendix} we provide simple calculations to demonstrate the compact nature of the millimeter emission.
This physical process would simultaneously explain a typical temperature of $>$60-{120} K (see above), which is significantly higher than that assumed for circumstellar disks associated with Class II YSOs heated by stellar radiation at a $\sim$100 au scale \citep[$\sim$20 K,][]{Williams11}.
%\\
%

{\it (B) Gravitational instability in the outer disk region} --- This physical process could be responsible for triggering the FUor outbursts (Section \ref{sec:intro}), and would fragment the outer disk region as shown by numerical simulations \citep[e.g.,][]{Vorobyov15b,Zhao18}. This would decrease the surface density in the bulk of the outer disk by forming compact, dense, and very optically thick clumps. As a result, this would decrease the millimeter continuum emission from the outer radii ($>$100 au), but a small fraction of leftover dust (of an order of $\sim$$10^{-6}$ $M_\sun$) would allow near-IR scattered light to be observed at these radii as shown by \citet{Liu16,Takami18}.
%\\
%

{\it (C) Grain growth and their inward drift} --- Numerical simulations by \citet{Vorobyov18} demonstrated that grain growth in the circumstellar disk can occur in the early stage of protostellar evolution ($t$$\sim$0.1 Myr), during which FUor outbursts can occur (Section \ref{sec:uni}). While small dust grains ($a < 1$ \micron) can be dynamically coupled with the gas content of the disk, large grains ($a > 1$ \micron) can lose the angular momentum via  friction with the gas component of the disk, and therefore gradually drift toward the star. \citet{Vorobyov18} demonstrated that, due to these physical processes, the fraction of millimeter- to micron-sized grains can be larger in the inner radii, in particular within $\sim$100 au of the star. This may be observed in millimeter emission as its compact intensity distribution, as millimeter-sized grains have a opacity significantly larger than micron-sized grains \citep[e.g.,][]{Wood02b,Dong12b}.
%\citet{Vorobyov18} demonstrated that the enhancement of mm-size grains, which are observable in millimeter continuum, can be observable at $r$$\sim$100 au in the disk, i.e., a spatial scale comparable to the millimeter emission observed for V900 Mon.

Among the above three physical processes, only (A) could simultaneously explain the warm and compact nature of the millimeter dust emission. Even so, processes (B) and/or (C) may also contribute to its compact nature.

If we attribute the continuum emission to the disk only, the observed spatial scale implies a disk inclination of up to 48\arcdeg~from the face-on. One may speculate that the disk has a larger inclination angle, close to edge-on, because (1) the CO intensity distribution of the extended outflow and the elongated envelope (Sections \ref{sec:results:vcm1}, \ref{sec:results:vcm2}, \ref{sec:discussion:envelope}); and (2) a relatively larger extinction despite a relatively low envelope mass (Section \ref{sec:intro}). This would require some contribution from a wind or a wind cavity to the observed flux.
The intense radiation from the inner disk would illuminate and heat the wind cavity close to the star, making thermal emission contribute to the observed flux. While the FUors are associated with a hot energetic ionized wind, as shown using optical spectrocopy \citep[e.g.,][]{Hartmann96,Reipurth12}, it would not be likely that free-free emission from a wind or jet significantly contributes to the observed flux, and therefore its angular scale. \citet{Liu17} executed a detailed analysis of millimeter and centimeter emission for FU Ori, and showed that free-free continuum emission from the ionized jet or the wind is $<$0.1 mJy at 230 GHz. The emission would be significantly fainter for V900 Mon, which has a distance $\sim$3 times farther than FU Ori .

%%%%%%%%%%%%%%%%%%%%%%%%%%%%%%%%%%%%%%%%
%%%%%%%%%%%%%%%%%%%%%%%%%%%%%%%%%%%%%%%%
%%%%%%%%%%%%%%%%%%%%%%%%%%%%%%%%%%%%%%%%
%% Section : Discussion (2)
%%%%%%%%%%%%%%%%%%%%%%%%%%%%%%%%%%%%%%%%
%%%%%%%%%%%%%%%%%%%%%%%%%%%%%%%%%%%%%%%%
%%%%%%%%%%%%%%%%%%%%%%%%%%%%%%%%%%%%%%%%

\section{Toward a Unified Scheme for Low-Mass Protostellar Evolution} \label{sec:uni}

The key issues for FUor outbursts are: (1) their triggering mechanism, and (2) whether {most} normal YSOs experience such accretion outbursts during their evolution (Section \ref{sec:intro}). A variety of circumstellar structures seen in near-IR scattered light suggest gravitational instabilities in disks as a triggering mechanism of the outbursts \citep{Liu16,Dong16,Takami18}. A combination of hydrodynamical simulations and radiative transfer calculations support this mechanism, and also the scenario that many YSOs experience similar accretion outbursts \citep{Dunham12,Dong16}. In this section we extend our discussion for the latter issue.

The evolutionary stages of normal low-mass YSOs are characterized by their infrared SEDs \citep[Class 0$\rightarrow$I$\rightarrow$II$\rightarrow$III;][]{Stahler05}. Some authors apply this SED classification for normal YSOs to FUors \citep[e.g.,][]{Green13,Gramajo14}. However, the SEDs of FUors during the outbursts may not indicate their evolutionary stages. This is because the repetition of the following processes would change the SED (Class I$\leftrightarrow$II) on a significantly shorter time scale than low-mass protostellar evolution: (1) an energetic FUor wind blowing away the circumstellar material responsible for a large mid- to far-IR excess; and (2) gas+dust infall from the outer to the inner envelope, recovering the circumstellar dust+gas responsible for the mid- to far-IR excess. We discuss this issue in detail in Section \ref{sec:uni:seds}.

In Sections \ref{sec:uni:mm1} and \ref{sec:uni:mm2} we compare observations of several FUors and FUor-like objects, and the envelope mass and the collimation of the extended CO outflow observed for a limited sample of normal Class 0-II YSOs. \citet{Arce06} observed the CO outflows and envelopes associated with Class 0, I, and II YSOs using millimeter interferometry. These authors showed that the envelope mass becomes smaller and the opening angle of the CO outflow becomes larger during the Class 0$\rightarrow$I$\rightarrow$II evolution. 
In Section \ref{sec:uni:mm1} we discuss the cases of V900 Mon and V346 Nor, for which the associated CO outflows and envelopes are similar to normal Class 0-I YSOs. These support the idea that {most} normal YSOs experience FUor outbursts during their evolution. In Section \ref{sec:uni:mm2} we discuss the cases for a few FUor-like objects (V883 Ori, HBC 494, V2775 Ori) for which the extended CO outflows may be widened by hot energetic FUor winds.

  %%%%%%%%%%%%%%%%%%%%%%%%%%%%%%%%%%%%%%%%
  %%%%%%%%%%%%%%%%%%%%%%%%%%%%%%%%%%%%%%%%
  %% Subsection : SEDs
  %%%%%%%%%%%%%%%%%%%%%%%%%%%%%%%%%%%%%%%%
  %%%%%%%%%%%%%%%%%%%%%%%%%%%%%%%%%%%%%%%%

\subsection{Possible Time Variation of Infrared SEDs} \label{sec:uni:seds}

\citet{Liu16,Takami18} executed near-IR imaging polarimetry of five classical FUors to observe scattered light in their circumstellar environment. Their polarized intensity distributions show a variety of morphologies with arms, tails or streams, spikes, and fragmented
distributions among the objects. The morphologies of these reflection nebulae differ significantly from many other normal YSOs. The authors attributed these structures to gravitationally unstable disks, trails of clump ejections, dust blown by a wind or a jet, and a stellar
companion.

\begin{figure}[ht!]
\epsscale{1}
\plotone{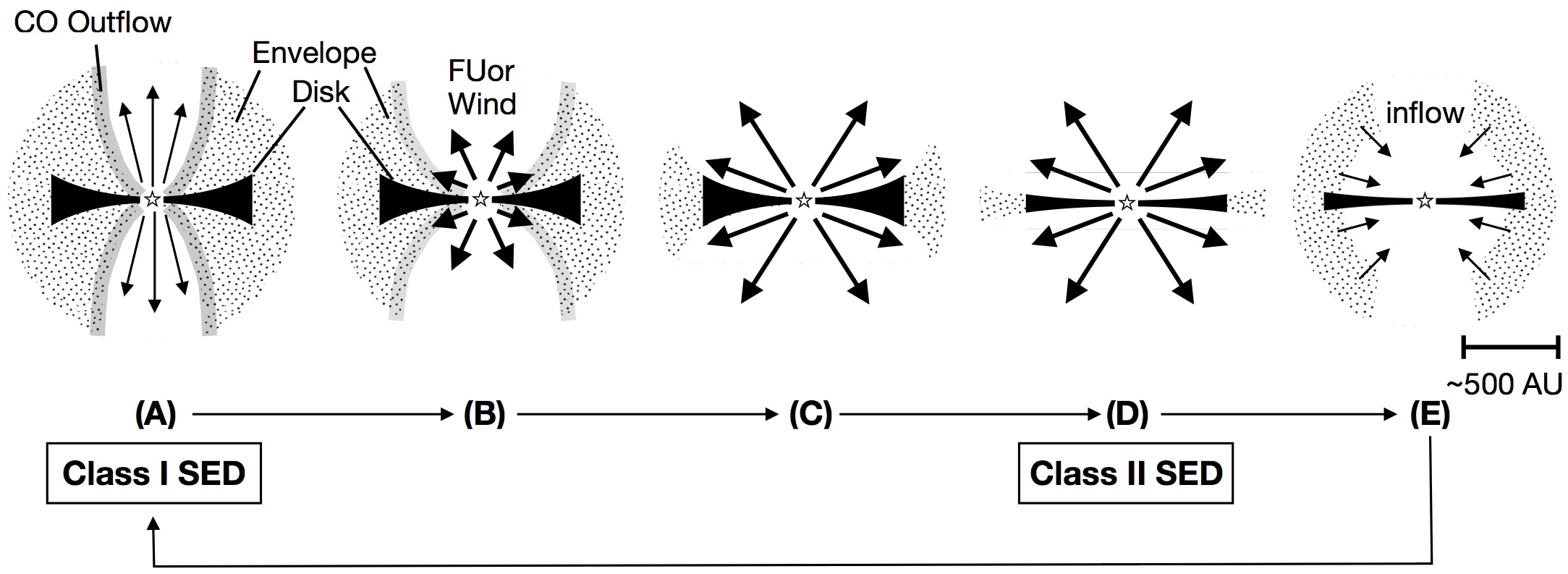}
\caption{The proposed sequence of the wind and infall associated with FUor outbursts. The arrows indicate the motion of gas flow due to a jet, a wind or infall. Adapted from \citet{Takami18}.
\label{fig:sequence}}
\end{figure}

Why do the FUors we observed look very different from each other in the near-IR? \citet{Takami18} proposed that YSOs follow the sequence summarized in Figure \ref{fig:sequence} with accretion outbursts and associated winds. A normal Class I YSO before the outburst is associated with a circumstellar disk, a circumstellar envelope and an extended CO outflow (A in the figure). The circumstellar envelope exists below the wall of the outflow cavity. The disk cannot be seen in the near-IR because it is embedded in an optically very thick circumstellar envelope. When an episodic energetic FUor wind emerges with the FUor outburst, it blows away the surface of the circumstellar disk and the inner region of the envelope at $\sim$1000 au (B-D in the figure). As a result, a variety of structures associated with the disk become visible in the near-IR, as for normal Class II YSOs. When the FUor wind stops, circumstellar material from the outer region infalls, quickly reassembling the disk and the inner envelope (E in the figure). The free-fall timescale is only $\sim$5000 years at 1000 au for a 1 $M_\sun$ central star. These processes would repeat during the Class I evolutionary phase \citep[i.e., a time scale of 4-5$\times$10$^5$ years, see][for a review]{Dunham14}, and gradually clear the outer envelope \citep[i.e. that observable using a single-dish {millimeter} telescope or in mid-IR silicate absorption; see e.g.,][]{Sandell01,Green06,Quanz07,Kospal17b} toward the Class II evolutionary phase discussed for normal YSOs \citep{Kospal17b}.

Given a typical luminosity of $\sim$200 $L_\sun$ for FUors, the mid- to far-IR SEDs would originate from the inner envelope or the outer disk ($r \lesssim 1000$ au) heated by radiation from the inner disk \citep{Zhu08} or the central star \citep{Elbakyan19}. \citet{Johnstone13} demonstrated that the thermal budget with the radiation cooling quickly reaches the equilibrium with a timescale of light-crossing time (i.e., $\sim$ 6 days for 1000 au). Therefore, the above precesses would change the infrared SED on a time scale significantly shorter than the Class I evolutionary phase.

  %%%%%%%%%%%%%%%%%%%%%%%%%%%%%%%%%%%%%%%%
  %%%%%%%%%%%%%%%%%%%%%%%%%%%%%%%%%%%%%%%%
  %% Subsection : V900 Mon, V346 Nor
  %%%%%%%%%%%%%%%%%%%%%%%%%%%%%%%%%%%%%%%%
  %%%%%%%%%%%%%%%%%%%%%%%%%%%%%%%%%%%%%%%%

\subsection{Implications of the Envelope and the Extended CO Outflows: V900 Mon and V346 Nor} \label{sec:uni:mm1}

Similar to many other YSOs, V900 Mon is associated with cold gas ($T$$<$50 K) in an extended molecular bipolar outflow and an envelope. Its spatial scale in our ALMA observations is at least $\sim$10$^4$ au (Sections \ref{sec:results} and \ref{sec:discussion}). Interaction with a hot energetic wind ($\Delta v \sim 200$ km s$^{-1}$), presumably associated with the FUor outburst, is limited to the region only up to $\sim$800 au (Section \ref{sec:discussion:wind}). Therefore, the intensity distribution for most parts of the extended CO outflow and the envelope has presumably remained the same since the pre-outburst phase, and we would be able to discuss the nature of the progenitor of this FUor.
%For V900 Mon, the envelope mass in the literature \citep[0.027 $M_\sun$][]{Gramajo14} and the CO outflow opening angle shown by our observations are similar to Class I YSOs, suggesting that the progenitor of this FUor object is a normal Class I YSO. 
For V900 Mon, the envelope mass in the literature \citep[0.027 $M_\sun$, ][]{Gramajo14} and the CO outflow opening angle shown by our observations {($\sim$70\arcdeg~and $\sim$120\arcdeg~at $r$$\sim$5000 au for the blueshifted and redshifted lobes, respectively)} are similar to Class I YSOs {\citep[e.g., 0.03-0.05 $M_\sun$; and 70\arcdeg-160\arcdeg~at $r$$\sim$5000 au; ][]{Arce06}}, suggesting that the progenitor of this FUor object is a normal Class I YSO. 

The millimeter observations of V346 Nor also show an envelope mass and extended CO outflow collimation similar to a normal YSO. \citet{Kospal17c} observed the CO outflows, the envelope, and the disk associated with this object using ALMA, in which the bipolar outflow extends over a $\sim$10$^4$ au scale. Its envelope mass \citep[0.3--1 $M_\sun$, ][]{Evans94,Sandell01,Kospal17b} and the outflow collimation are similar to Class 0 YSOs rather than Class I YSOs as shown by \citet{Arce06}.

As with V900 Mon, V346 Nor does not show clear evidence for interaction between the hot FUor wind and the extended CO outflow and the envelope. This star went into eruption some time between 1976 and 1980, and showed a rapid fading in 2010-11, then brightened again \citep[see][for a summary]{Kospal17c}. If the energetic wind emerged at the onset of the outburst in 1976--1980, the wind would reach up to $\sim$5000 au, adopting a velocity similar to the other FUors \citep[up to $\sim$600 km s$^{-1}$;][]{Audard14}.
However, the CO observations made by \citet{Kospal17c} do not show discontinuous intensity distributions between the inside and the outside of this radii up to $\sim$10$^4$ scales. Therefore, we assume that we can discuss the progenitor based on the associated envelope and extended CO outflows.

  %%%%%%%%%%%%%%%%%%%%%%%%%%%%%%%%%%%%%%%%
  %%%%%%%%%%%%%%%%%%%%%%%%%%%%%%%%%%%%%%%%
  %% Subsection : Another few cases
  %%%%%%%%%%%%%%%%%%%%%%%%%%%%%%%%%%%%%%%%
  %%%%%%%%%%%%%%%%%%%%%%%%%%%%%%%%%%%%%%%%

\subsection{Cases of V883 Ori, HBC 494, V2775 Ori} \label{sec:uni:mm2}

These FUor-like objects are associated with an envelope with 0.1-0.4 $M_\sun$\citep{Sandell01,Caratti11,Gramajo14,Kospal17b}, comparable to those of Class 0 YSOs \citep[$\sim$0.2 $M_\sun$][]{Arce06}. However, the morphologies of their CO outflows observed using ALMA differ significantly from those of Class 0 protostars. \citet{Ruiz17_V883Ori,Ruiz17_HBC494} showed that the outflows associated with HBC 494 and V883 Ori have opening angles significantly larger than these of Class 0 YSOs, similar to Class II YSOs observed by \citet{Arce06}. \citet{Zurlo17} showed that the star is associated with a pair of blueshifted and redshifted rings or shells in the CO emission, that can be explained by a bipolar outflow but are not usually observed toward normal Class 0-II YSOs.

The outflow morphologies inconsistent with the Class 0 phase (i.e., these suggested by the envelope mass) can be explained if their outflows are due to interactions between the ambient gas and a hot energetic FUor wind as suggested by \citet{Ruiz17_V883Ori} for V883 Ori. However, \citet{Zurlo17} pointed out that the latest outbursts associated with V2775 Ori cannot explain the observed CO outflow at 4000--8000 au considering the dates for observations and a typical outflow velocity. In this context, the CO outflows associated with V2775 Ori may have been driven by winds associated with the previous outbursts.

Why would we see interaction with the FUor winds associated with the previous outbursts for V2775 Ori but not for V900 Mon or V346 Nor?
The time intervals between the FUor outbursts and the associated winds may span a wide range of {scales for} different stars, and even for the same stars. \citet{Vorobyov15b,Vorobyov18_jet} executed numerical hydrodynamics simulations with gravitationally unstable disks and reproduced outbursts with time intervals of 30--10$^5$ yrs. According to their simulations, compact infalling fragments trigger isolated outbursts with time intervals between the outbursts of $>$1000 yrs if the tidal {torques do} not disintegrate the fragments. If a fragment starts disintegrating when it is migrating to the inner disk, it will cause a cluster of outbursts with time intervals of $<$1000 yrs. In these contexts, the time intervals of FUor outbursts associated with individual objects would highly depend on the disk conditions and hence on the initial conditions in their parental prestellar cores and their evolutionary status.

%The frequency of the FUor outbursts (and therefore emergence of an associated wind) for each YSO would be very low considering the fact that only about a dozen FUors and another dozen FUor-like objects have been discovered to date \citep{Audard14}, and considering the presence of a number of known YSOs at similar distances. In this context, V2775 Ori may be a peculiar FUor, or another category of outbursting YSO, which is associated with more frequent emergence of the accretion outbursts and {\bf hot} wind. The 1.1-2.5 $\micron$ spectrum obtained by \citet{Caratti11} shows absorption features similar to FUors, but with Br $\gamma$ emission (2.17 $\micron$) as in EXors, another category of bursting stars with variation on a significantly shorter time scale than FUors \citep[][]{Hartmann96,Audard14}. We note that HBC 494 may also be a peculiar FUor, as its $K$-band (1.9-2.4 $\micron$) spectrum does not show any emission or absorption features, in contrast to those of the other FUors that show absorption associated with the disk atmosphere \citep[e.g.,][]{Hartmann96,Reipurth97}.

%%%%%%%%%%%%%%%%%%%%%%%%%%%%%%%%%%%%%%%%
%%%%%%%%%%%%%%%%%%%%%%%%%%%%%%%%%%%%%%%%
%%%%%%%%%%%%%%%%%%%%%%%%%%%%%%%%%%%%%%%%
%% Section : Conclusions
%%%%%%%%%%%%%%%%%%%%%%%%%%%%%%%%%%%%%%%%
%%%%%%%%%%%%%%%%%%%%%%%%%%%%%%%%%%%%%%%%
%%%%%%%%%%%%%%%%%%%%%%%%%%%%%%%%%%%%%%%%

\section{Conclusions} \label{sec:conclusions}

We observed $^{12}$CO, $^{13}$CO, and C$^{18}$O $J$=2--1 lines and the 230 GHz  continuum for the FU Ori-type object V900 Mon ($d$$\sim$1.5 kpc) using ALMA, with a 0\farcs2$\times$0\farcs15 resolution. The $^{12}$CO maps show the presence of an extended molecular bipolar outflow in the east-west direction at a $\sim$10$^4$ au scale. The $^{13}$CO maps show compressed gas at the cavity wall. The C$^{18}$O maps show the presence of rotating envelope across the jet axis extending over a $\sim$10$^4$ au scale. 

Previous optical spectroscopy shows the presence of a hot energetic FUor wind at the star, with $v$$\sim$200 km s$^{-1}$. This wind may be responsible for high-velocity ($\Delta v$ up to $\sim$5 km s$^{-1}$) $^{12}$CO emission and a possible cavity in the C$^{18}$O emission within 0\farcs5 ($\sim$800 au) of the star. We do not have any clear evidence for this wind interacting with molecular gas at larger distances. This suggests that the most parts of the extended CO outflows and the envelope are the same as before the onset of the outburst.

The envelope mass and the CO outflow collimation of V900 Mon suggest that the progenitor of this FUor is a normal Class I YSO. Similarly, the progenitor of another FUor, V436 Nor, may be a normal Class 0 YSOs. These results are consistent with the idea that FUor outbursts occur for {most} YSOs. In contrast, extended CO outflows associated with some FUor-like stars (V883 Ori, HBC 494, V2775 Ori) may be driven or widened by a hot energetic wind which is often observed toward FUors.

As with other FUor and FUor-like stars, V900 Mon is associated with compact continuum millimeter emission. We measured a FWHM angular scale of 0\farcs06 ($\sim$90 au) using a two-dimensional Gaussian fitting for the millimeter emission and the beam, respectively. The emission is associated with a dusty circumstellar disk, plus possible emission from a wind or a wind cavity near the base. The temperature of the region seems to be significantly higher than usually assumed for a disk associated with normal YSOs.

The observed spatial scale of probable disk emission is significantly smaller than the disks associated with some FUors and that seen in the near-IR. The warm and compact nature of the disk continuum emission could be explained with viscous heating of the disk, although gravitational fragmentation in the outer disk and/or a combination of grain growth and their inward drift may also contribute to its compact nature.

%To investigate the compact nature of the emission, we compared the observed spatial extension using a conventional viscous accretion model, which explains the observed luminosity and the millimeter fluxes. The model provides an angular scale larger than the observations. The discrepancy can be overcome if the outer disk region is fragmented due to gravitational instabilities, whose signatures are seen in some near-IR observations.

%% If you wish to include an acknowledgments section in your paper,
%% separate it off from the body of the text using the \acknowledgments
%% command.
\acknowledgments

This project has received funding from the European Research Council (ERC) under the European Union's Horizon 2020 research and innovation programme under grant agreement No 716155 (SACCRED).
MT and TSC are supported from Ministry of Science and Technology (MoST) of Taiwan (Grant
No. 106-2119-M-001-026-MY3). 
MT and HBL are supported from MoST of Taiwan 108-2923-M-001-006-MY3 for the Taiwanese-Russian collaboration project.
EIV acknowledges support from the Russian Foundation for Basic Research (RFBR), Russian-Taiwanese project \#19-52-52011. 
This paper makes use of the following ALMA data:
ADS/JAO.ALMA \#2016.1.00209.S and \#2017.1.00451.S. ALMA is a partnership of ESO (representing
its member states), NSF (USA) and NINS (Japan),
together with NRC (Canada), MoST and ASIAA (Taiwan),
and KASI (Republic of Korea), in cooperation with the Republic
of Chile. The Joint ALMA Observatory is operated
by ESO, AUI/NRAO and NAOJ. This work has made use of
data from the European Space Agency (ESA) mission Gaia
(https://www.cosmos.esa.int/gaia), processed by the Gaia
Data Processing and Analysis Consortium (DPAC, https://www.cosmos.esa.int/web/gaia/dpac/consortium). Funding
for the DPAC has been provided by national institutions, in
particular the institutions participating in the Gaia Multilateral
Agreement.
This research made use of the
Simbad database operated at CDS, Strasbourg, France, and the
NASA's Astrophysics Data System Abstract Service.

%% To help institutions obtain information on the effectiveness of their 
%% telescopes the AAS Journals has created a group of keywords for telescope 
%% facilities.
%
%% Following the acknowledgments section, use the following syntax and the
%% \facility{} or \facilities{} macros to list the keywords of facilities used 
%% in the research for the paper.  Each keyword is check against the master 
%% list during copy editing.  Individual instruments can be provided in 
%% parentheses, after the keyword, but they are not verified.

\vspace{5mm}
\facilities{ALMA}

%% Similar to \facility{}, there is the optional \software command to allow 
%% authors a place to specify which programs were used during the creation of 
%% the manusscript. Authors should list each code and include either a
%% citation or url to the code inside ()s when available.

\software{CASA\citep{McMullin07}, numpy {\citep{numpy}}, scipy {\citep{scipy}}}

%% Appendix material should be preceded with a single \appendix command.
%% There should be a \section command for each appendix. Mark appendix
%% subsections with the same markup you use in the main body of the paper.

%% Each Appendix (indicated with \section) will be lettered A, B, C, etc.
%% The equation counter will reset when it encounters the \appendix
%% command and will number appendix equations (A1), (A2), etc. The
%% Figure and Table counter will not reset.

\appendix

%%%%%%%%%%%%%%%%%%%%%%%%%%%%%%%%%%%%%%%%
%%%%%%%%%%%%%%%%%%%%%%%%%%%%%%%%%%%%%%%%
%%%%%%%%%%%%%%%%%%%%%%%%%%%%%%%%%%%%%%%%
%% Appendix A 
%%%%%%%%%%%%%%%%%%%%%%%%%%%%%%%%%%%%%%%%
%%%%%%%%%%%%%%%%%%%%%%%%%%%%%%%%%%%%%%%%
%%%%%%%%%%%%%%%%%%%%%%%%%%%%%%%%%%%%%%%%

\section{Millimeter Continuum Emission from a Viscous Disk} \label{sec:appendix}

We use a simple conventional flat disk model, used extensively for FUors over many years \citep[e.g.,][]{Calvet91b,Zhu07,Zhu08}, to demonstrate the compact nature of the observed millimeter emission. According to this model, the accretion luminosity from the disk and the temperature distribution are described as follows \citep[e.g.,][]{Zhu07}:
\begin{eqnarray}
L_{\mathrm acc} &=&\frac{GM_*\dot{M}}{2R_i}, \\
T_{\rm eff}^4 (r) &=& \frac{3GM_*\dot{M}}{8 \pi \sigma r^3} \left[ 1- \left( \frac{R_i}{r} \right)^{1/2} \right],
\end{eqnarray}
where $L_{\mathrm acc}$ is the accretion luminosity;
$G$ is the gravitational constant;
$M_*$ is the stellar mass;
$\dot{M}$ is the mass accretion rate;
$R_i$ is the disk inner radius (= the stellar radius);
$T_{\rm eff}(r)$ is the effective temperature at the disk radius $r$;
and $\sigma$ is the Stefan-Boltzmann constant.
%and $r$ is radius in the disk.
From these equations we derive:
\begin{eqnarray}
%\dot{M}  &=&\frac{2R_iL}{GM_*}, \\
T_{\rm eff} (r) &=& \left\{ \frac{3R_i L_{\mathrm acc}}{4 \pi \sigma r^3} \left[ 1- \left( \frac{R_i}{r} \right)^{1/2} \right] \right\}^{1/4},
\end{eqnarray}
Equation (A3) indicates that we can calculate the temperature distribution with the given disk radius, the inner disk radius and the accretion luminosity from the disk, without measuring/assuming the stellar mass and/or the mass accretion rate.

The intensity from each part of the disk would be described as:
\begin{eqnarray}
I_\nu(r) &=& B_\nu(r) (1-e^{-\tau (r)}), \\
B_\nu(r) &=& \frac{2h\nu^3}{c^2} \frac{1}{{\mathrm exp}[\frac{h\nu}{kT_{\mathrm eff}(r)}]-1},
\end{eqnarray}
where $I(r)$ is the intensity; 
$B(r)$ is the blackbody radiation;
$h$ is the Planck constant;
$\nu$ is the frequency of the observations;
$c$ is light speed;
and $k$ is the Boltzmann constant.
The parameter $\tau (r)$ is optical thickness, which is described as $\kappa_{ext} \Sigma (r)$(cos $i$)$^{-1}$, where $\kappa_{ext}$ is the dust opacity; $\Sigma (r)$ is dust surface density; and $i$ is the disk inclination angle. Here we adopt $\Sigma (r) \propto r^{-1}$ based on previous observations of disks for Class II YSOs \citep[e.g.,][]{Williams11,Takami13}. Therefore:
\begin{equation}
\tau (r) = \kappa_{ext} \Sigma_0 r^{-1} (\mathrm{cos}~i)^{-1}.
\end{equation}
We discuss submillimeter emission from dust continuum, therefore we assume:
\begin{equation}
\tau(r) = 0 \rm{~~~for~~} \it T > T_{\rm sub},
\end{equation}
where $T_{\rm sub}$ is the sublimation temperature of dust grains.  We note that if the maximum temperature in the disk is above $T_{\rm sub}$,  Equations (A2) and (A3) provide two radii for the sublimation temperature, one comparable to the stellar radius $R_i$, the other significantly larger. This is because Equations (A2) and (A3) provide the maximum temperature $T_{\mathrm max}$ at $r$=1.36 $R_i$, and the temperature monotonically decreases inward and outward, respectively \citep{Zhu07}. However, \citet{Zhu07} pointed out that these equations provide an unphysical temperature distribution at $r$$<$1.36 $R_i$. Therefore, we only use the outer sublimation radius (hereafter $r_{\mathrm sub}$), and assume $I_\nu(r)=0$ and $\tau(r)=0$ for $r < r_{\mathrm sub}$.

The spatially integrated flux is described as:
\begin{equation}
F_\nu = \int_{R_i}^{r_{out}} \frac{2 \pi r I(r)~\mathrm{cos}~i}{d^2} dr = \frac{2 \pi~\mathrm{cos}~i}{d^2} \int_{r_{sub}}^{r_{out}} r I(r) dr,
\end{equation}
where $r_{out}$ is the outer radius of the disk; and $d$ is the distance to the target. In this equation we replace the inner disk radius $R_i$ with the sublimation radius $r_{\mathrm sub}$ because the latter should be significantly larger than the former.
%Note that the sublimation radius $r_{\mathrm sub}$ is significantly larger than the inner disk radius $R_i$.
This is corroborated with the fact that the FUors are associated with optical and near-IR emission from the inner disk with a spectral type of K-M or earlier, indicating that the temperature of the inner disk region exceeds over $T \gtrsim 3000$ K \citep{Hartmann96,Audard14}.

We can calculate the flux $F_\nu$ using Equations (A3)-(A8),
as a function of the accretion luminosity from the disk $L_{\mathrm acc}$,
the frequency of the observations $\nu$,
the dust opacity $\kappa_{ext}$,
the constant to scale the surface density $\Sigma_0$,
the disk inclination angle $i$,
the inner radius $R_i$ % or the sublimation radius,
the outer radius $r_{out}$,
and the distance $d$.
With the given $L_{\mathrm acc}$, $\nu$, $\kappa_{ext}$, $i$, $R_i$, $r_{\mathrm out}$ and $d$, the flux $F_\nu$ is a function of $\Sigma_0$ only. We adjusted $\Sigma_0$ to match the flux to the observations using {\tt scipy.interpolate.interp1d}.

\begin{figure}[ht!]
\epsscale{1}
\plotone{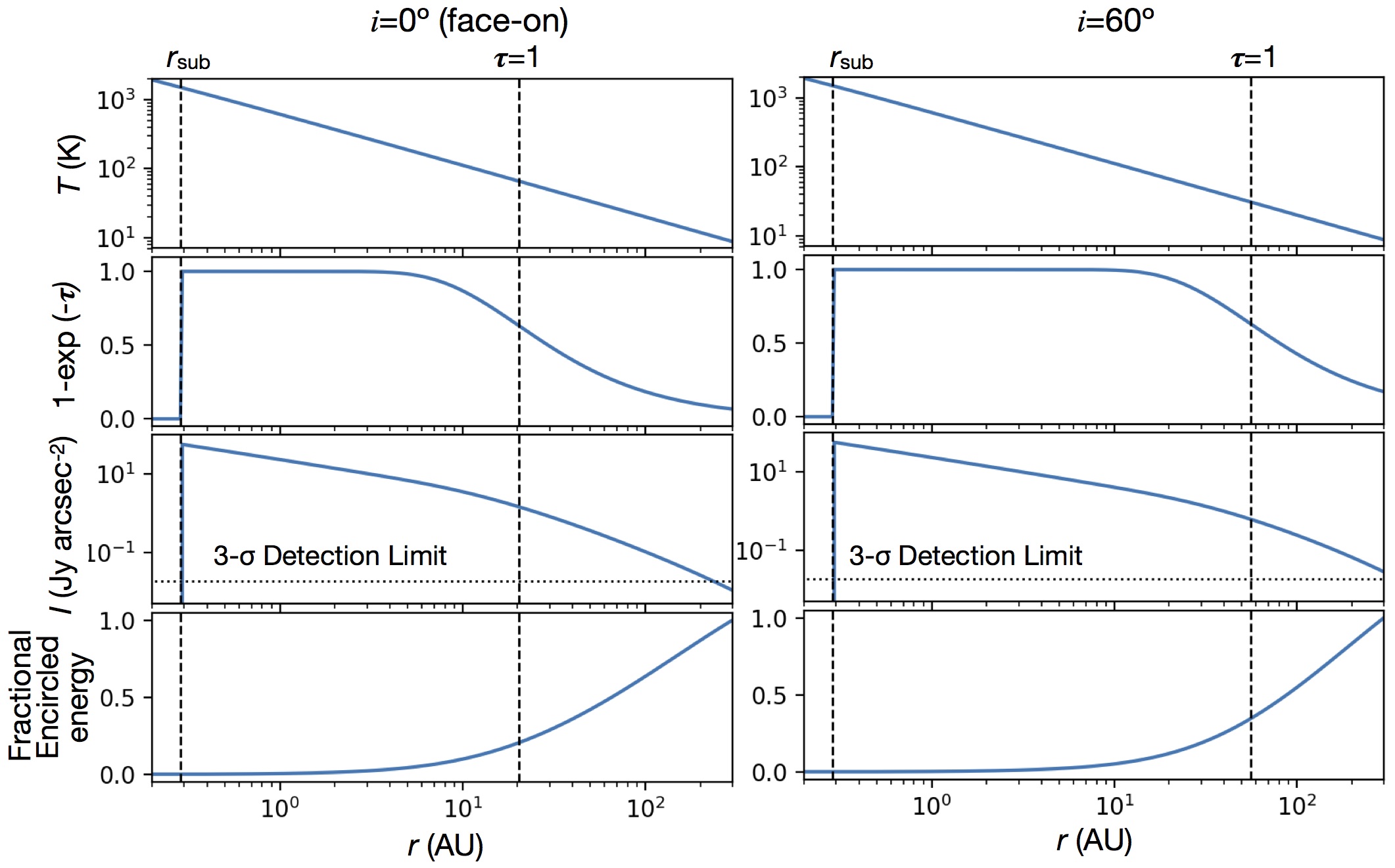}
\caption{Physical parameters of the viscous disk model (see text for details). The left and right panels are for the disk inclination angles of 0\arcdeg~(face-on view) and 60\arcdeg, respectively. The vertical dashed lines show the dust sublimation radius and the radius for optical thickness $\tau$=1. The horizontal dotted line in the third box shows the detection limit of our ALMA observations.
\label{fig:viscous_disk}}
\end{figure}

The upper part of Table \ref{tab:model} summarizes the parameters we used for the simulations. We tentatively set the disk accretion luminosity $L_{\mathrm acc}$, a dust opacity $\kappa_{\mathrm ext}$, and the outer disk radius $r_{\mathrm out}$ as follows:
\begin{itemize}
\item {\it $L_{\mathrm acc}$} --- We assume 100 $L_\sun$, about half of the observed bolometric luminosity ($\sim$200 $L_\sun$; Section \ref{sec:intro} and Table \ref{tab:ALMA_FUors}) for the following reason. The primary energy sources of the observed bolometric luminosity would be accretion heating in the disk \citep[see e.g.,][for a review]{Hartmann96} and at the central star \citep[][]{Baraffe12,Elbakyan19}. The fraction of these disk/stellar accretion luminosities is uncertain, and also time-variable \citep[][]{Elbakyan19}. Fortunately, our calculations depend on the assumed accretion luminosity only marginally as seen in Equation (A3).

\item {\it $\kappa_{\mathrm ext}$} --- We assume 0.2 m$^2$ kg$^{-1}$ following \cite{Beckwith90}. This value is also time-variable with grain growth \citep{Vorobyov18}. A larger fraction of millimeter-size grains would provide a lower dust opacity \citep[e.g.,][]{Dong12b}, and therefore a large surface density and a total disk mass based on Equation (A7).

\item
{\it $r_{\mathrm out}$} --- We assume 300 au, i.e., comparable to the area where we measure the flux and the FWHM of the continuum emission. In reality, the disk emission may extend beyond this radius below the detection limit (Figure \ref{fig:viscous_disk}). 

\end{itemize} 

  %%%%%%%%%%%%%%%%%%%%%%%%%%%%%%%%%%%%%%%%
  %% Table: Model parameters
  %%%%%%%%%%%%%%%%%%%%%%%%%%%%%%%%%%%%%%%%

\begin{deluxetable}{lcc}[b!]
\tablecaption{Model Parameters \label{tab:model}}
\tablecolumns{3}
%\tablenum{2}
\tablewidth{0pt}
\tablehead{
\colhead{Initial Paramerters} 
%\multicolumn{3}{c}{Initial Parameters}
%\colhead{Paramerter} &
%colhead{Value} 
}
\startdata
Accretion luminosity from the disk ($L_{\mathrm acc}$)	& \multicolumn{2}{c}{100 $L_\sun$\tablenotemark{a}}		\\
Frequency ($\nu$)					& \multicolumn{2}{c}{230 GHz}		 				\\
Total Flux ($F_\nu$)					& \multicolumn{2}{c}{9.0 mJy}		 					\\
Dust opacity ($\kappa_{\mathrm ext}$)	& \multicolumn{2}{c}{0.2 m$^2$ kg$^{-1}$\tablenotemark{b}}	\\
Inner disk radius ($R_{\mathrm i}$) 		& \multicolumn{2}{c}{5 $R_\sun$\tablenotemark{c}}			 \\
Outer disk radius ($r_{\mathrm out}$)	& \multicolumn{2}{c}{300 au}		\\
Distance							& \multicolumn{2}{c}{1500 pc}		\\
Sublimation temperature $T_{\mathrm sub}$ & \multicolumn{2}{c}{1500 K\tablenotemark{d}}	\\ \tableline
%%%
Derived Parameters & $i$=0\arcdeg & $i$=60\arcdeg\\ \tableline
%%%
Dust surface density $\Sigma$ at $r$=1 au (kg m$^{-2}$)	& 1.0$\times$10$^2$	& 1.4$\times$10$^2$	\\
Sublimation radius $r_{\mathrm sub}$ (au)			& 0.29	& 0.29 \\
Radius for $\tau$=1 (au)							& 20.3 	& 55.8 \\
Total dust mass ($M_\sun$)		& 2.2$\times$10$^{-3}$  & 3.0$\times$10$^{-3}$  
\enddata
\tablenotetext{a}{See text}
\tablenotetext{b}{\cite{Beckwith90}}
\tablenotetext{c}{\cite{Zhu07,Gramajo14}}
\tablenotetext{d}{\cite{Lodders03}}
%\tablenotetext{z}{\bf EV: if this is dust density, then it corresponds to a gas density of about 30 g cm$^{-2}$ for a typical 1:100 dust-to-gas ratio. A typical MRI outburst is triggered at 10000 g cm$^{-2}$. This means that the inner disk has already been significantly depleted in material due to the outburst. So, this is indeed an evolved stage of the outburst.}
\end{deluxetable}

\begin{figure}[ht!]
\epsscale{1}
\plotone{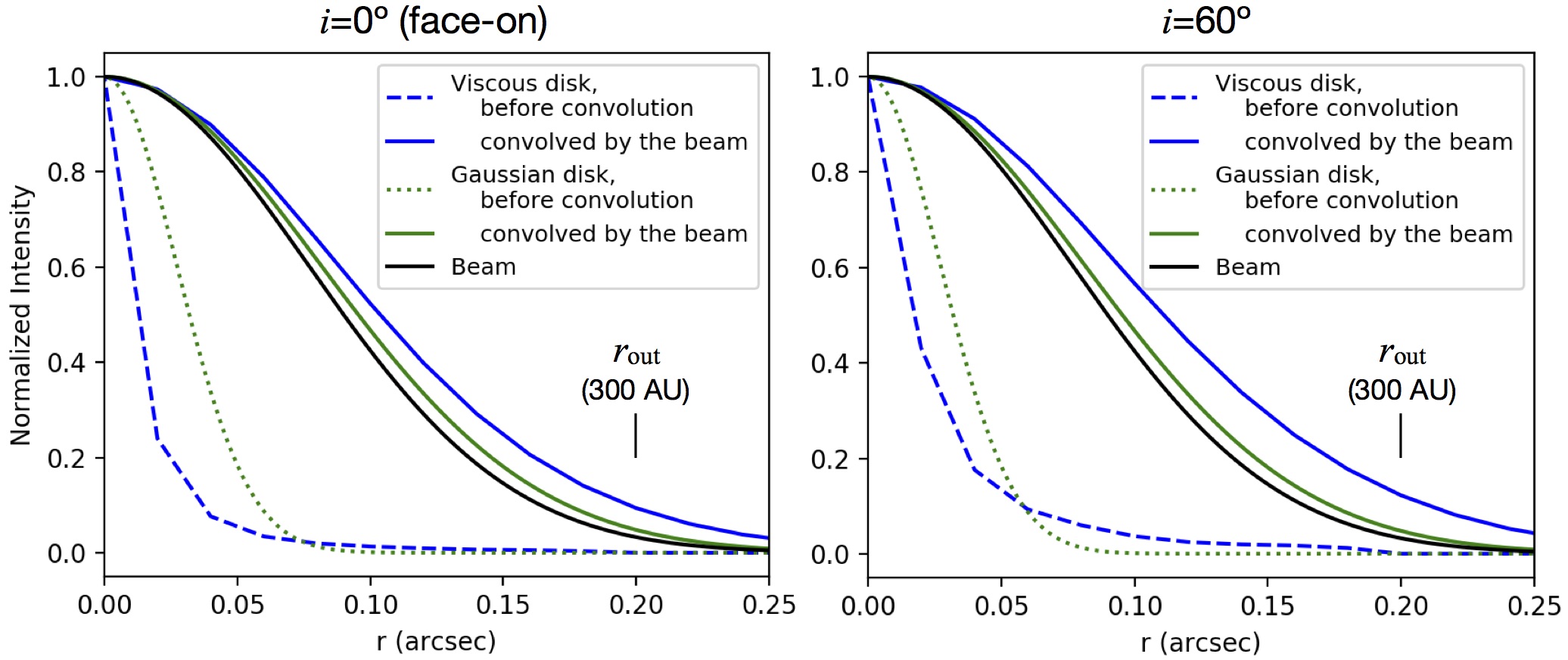}
\caption{The spatial distribution of the viscous and Gaussian disk models and their pseudo-observations with a Gaussian with a FWHM=0\farcs18 beam. The left and right panels are for the disk inclination angles of 0\arcdeg~(face-on view) and 60\arcdeg, respectively. We use a 0\farcs02 pixel sampling (i.e., the same as the observed data) for the original and convolved distributions of the viscous disk models. 
\label{fig:model_vs_obs}}
\end{figure}

Figure \ref{fig:viscous_disk} and the lower part of Table \ref{tab:model} show the modeled results for the disk inclination angles of 0\arcdeg~(face-on view) and 60\arcdeg. To compare the intensity distribution with the observations, we conducted monochromatic Monte-Carlo simulations using the intensity distribution in Figure \ref{fig:viscous_disk}, and convolved with pseudo-beam of the observations, i.e., a two-dimensional Gaussian with a FWHM of 0\farcs18. Similarly, we convolved the ``Gaussian disk" used in Section  \ref{sec:results:continuum} (0\farcs064 FWHM) using the same beam to represent our ALMA observations in that section. Their profiles are shown in Figure \ref{fig:model_vs_obs}. We used 10$^5$ photons to obtain the image of the viscous disk at each disk inclination angle.

  %%%%%%%%%%%%%%%%%%%%%%%%%%%%%%%%%%%%%%%%
  %% Figure: Property of the viscous disk model
  %%%%%%%%%%%%%%%%%%%%%%%%%%%%%%%%%%%%%%%%

  %%%%%%%%%%%%%%%%%%%%%%%%%%%%%%%%%%%%%%%%
  %% Figure: Comparison between the model & observation
  %%%%%%%%%%%%%%%%%%%%%%%%%%%%%%%%%%%%%%%%

As shown in Figure \ref{fig:model_vs_obs}, the viscous disk models provide intensity distributions highly concentrated at the center. In contrast, the outer disk region (like $r$$>$50-100 au) also contributes to the total flux as significantly as the inner disk (Figure \ref{fig:viscous_disk}), and therefore to the observed intensity distribution after the convolution. In Figure \ref{fig:model_vs_obs} both the viscous and Gaussian disk models show marginal extension beyond the beam of the observations, but the former is slightly larger than the latter. 
The viscous disk models yield a FWHM of the observations of 0\farcs21 and 0\farcs22, respectively. These are 9 \% and 16 \% larger than that with the Gaussian disk model, i.e., that represents the ALMA observations in Section 3.1.

More realistic models, which are beyond the scope of the paper, require the following treatments: (1) vertical temperature gradients in the disk \citep[e.g.,][]{Calvet91a,Calvet91b,Vorobyov14}; (2) scattering of the millimeter radiation in the disk, that would degrade the emission from the inner disk region \citep[e.g.,][]{Liu19a}; (3) illumination of the outer disk by the inner disk or the central star \citep[e.g.,][]{Zhu07,Liu16,Dong16,Cieza18}; (4) low Planck opacities in the outer disk, which may make Equation (A2) invalid.

\end{document}